\documentclass[fleqn,usenatbib]{mnras}

\usepackage[T1]{fontenc}

\DeclareRobustCommand{\VAN}[3]{#2}
\let\VANthebibliography\thebibliography
\def\thebibliography{\DeclareRobustCommand{\VAN}[3]{##3}\VANthebibliography}

\usepackage{graphicx}	
\usepackage{amsmath}	
\usepackage{amssymb}	
\usepackage{multirow}
\usepackage{float}
\usepackage{xcolor}

\definecolor{szary}{gray}{0.8}

\title[47 Kepler contact binaries]{The light curve intrinsic variability in 47 Kepler contact binary stars}

\author[B. Debski]{
B. Debski\thanks{E-mail: b.debski@oa.uj.edu.pl}
\\
Astronomical Observatory, Jagiellonian University, Orla 171, 30-244 Krakow, Poland
}

\date{Accepted XXX. Received YYY; in original form ZZZ}

\pubyear{2021}

\usepackage{newtxtext,newtxmath}

\begin{document}
\label{firstpage}
\pagerange{\pageref{firstpage}--\pageref{lastpage}}
\maketitle

\begin{abstract}
This work studies the significance of the light curve intrinsic variability in the numerical modeling of contact binaries. Using synthetic light curves we are showing that the starspot-based intrinsic variability increases the apparent mass ratio by $\Delta q=5\%$. For systems with orbital period $P>0.3$\,d the effect of intrinsic variability averaged over long time cancels each other out with the Kepler Mission-like phase smearing. Further, we analyse 47 totally eclipsing Kepler Mission contact binaries. We found a sharp cutoff of the intrinsic variability at $P=0.45$\,d. With the light curve numerical modeling and observational relations we derive physical parameters of the 47 systems. At least 53\% of binaries have a possible third companion. 21 binaries show the O'Connell effect in the averaged phase curve. 19 of them have a primary maximum lower than the secondary, suggesting a stationary dark region on the trailing side. Using the $P=0.45$\,d cutoff we propose a new approach on the Period-Color relation. The only parameter correlating with the magnitude of the intrinsic variability is the apparent effective temperature ratio. We conclude that instead of describing the system parameters, the A/W-subtype division should be applicable only to the light curves, as a tentative phenomenon.  
\end{abstract}




\section{Introduction}

Contact binary stars of the W UMa-type are one of the most interesting examples of close binaries. According to their canonical model \citep{lucy68a,lucy68b}, they consist of two main sequence stars, which share a common convective envelope. 
Following the Roche geometry, the contact configuration means that both components exceed their Roche lobes and the common surface of a binary lies somewhere between the inner and outer critical Lagrangian equipotential surfaces. The Roche geometry itself is controlled via one parameter only: the mass ratio of the binary components \citep[see][]{kopal,hilditch}. With a shared surface and only one parameter driving the geometry, the process of numerical modeling of the light curves of contact binaries is relatively straightforward. As a direct consequence of that, there is large number of publications with contact binary models \citep[such as][to mention some]{2003A&A...412..465K,2006AJ....131.2986P,2004A&A...426.1001C}. Great many of them utilizing the Wilson-Devinney code \citep[][]{1971ApJ...166..605W} or its adaptations (such as Phoebe 1.0\footnote{\url{http://phoebe-project.org/1.0}}).
Such systems are proven to last for a very long time, reaching even a few billions of years, before final merging \citep[][]{2003ASPC..293...76W, 2008MNRAS.390.1577G}. Their slow, quasi-stable evolution is explained well by the Thermal Relaxation Oscillations theory \citep[][]{1976ApJ...205..217F,1976ApJ...205..208L}. 

Despite a long history of observations, rich bibliography and a multitude of case study light curve models, the interiors of contact binaries are still rather poorly understood. Even the surface itself remains a conundrum. For example, the common convective envelope should have a, more or less, uniform effective temperature. This follows, the light curves of contact binaries should have primary and secondary minima of nearly equal depths. Observations show that this holds even for the most extreme mass ratios \citep[such as $q=0.044$, see][]{2018RNAAS...2...13S}. While such situation is true in general, the light curves of contact binaries tend to exhibit significant intrinsic variations. The same intrinsic variations are causing the time dependent short-scale minima timing variations \citep[][]{tran2013} and variations of the O'Connell effect \citep[see e.g.][]{1951PRCO....2...85O,debski20}. This phenomenon is being explained collectively with the presence of starspots \citep[the 'subluminous regions' in][]{1965VeBam..27...36B}. \citet{mullan75} pointed out that such phenomena can be of magnetic origin. But the problem with the starspot-based intrinsic variability is that in order to make the starspot migrate, a differential rotation should be introduced. How to merge the necessity of a differential rotation with the canonical model of contact binaries is not fully resolved yet. Nonetheless, the existence of starspots leads to some interesting scenarios, such as estimating the thickness of the convective zone, basing on the spot coverage \citep[see, e.g.][]{hendry,senavci}. Starspots could be therefore used as a probing tool for introductory studies of the interiors of contact binaries. 

The general issue with that approach lies within the reliability of the spot parameters. Usually, spots are introduced in the numerical modeling as a circular region on one of the stars. They are defined by their position (longitude and co-latitude), relative temperature, and radius \citep[e.g. in the Wilson-Devinney code as of][]{1979ApJ...234.1054W}. Out of those four parameters only the longitudinal position can be recovered reliably \citep[see, e.g.][]{1985A&A...152...25V}. The size of a spot and its distance from the stellar pole are so highly entangled, that it is practiced by many to just fix the position of the spot on the equator \citep[e.g.][]{2010AJ....140..215Z}. Such an approach, while greatly reduces the computational time, is obviously based on no physical premises. In the end, the resulting model is not quite useful when it comes to analyze the spot itself.

With this article we begin our study of the light curve intrinsic variability understood as a direct consequence of the photospheric phenomena. Our general approach is to treat the intrinsic variability as a tool for testing the starspot migration models. We study the 'evolution of the light curve', as shown in the preliminary works in \citet{debski14, debski15} and \citet{debski20}. Such initiative was possible only with the release of a multitude of the long time-base, precise light curves provided by the Kepler mission \citep[][]{borucki}. Our methods and the resulting catalog of light curve intrinsic variability in more than 1200 objects are to be described elsewhere\footnote{A preliminary preview of the catalog is possible at\\ \url{http://bade.space/lcma/}}. This work is focused on building a primary set of contact binaries, with well established system parameters. One of the main parts of this endeavour is the analysis of how the intrinsic variability can influence the results of the numerical modeling in case of contact binaries. Hence, we begin this work with an introduction of how the averaging of the intrinsic variability affect a phase-folded light curve. In the next step, we analyse the intrinsic variability averaging combined with the phase smearing \citep[][]{zola2017}. The numerical modeling scheme and resulting models are shown in Section~\ref{sec:modeling}. In the final part we explore the physical parameters of the objects and discuss the results in the light of recent literature.

\section{Sample selection}
\label{sec:sample}

We composed the set of contact binary candidates using the Kepler Eclipsing Binary Catalogue\footnote{\url{http://keplerebs.villanova.edu/}, accessed in February 2019} \citep[hereafter KEBC,][]{kebc}. The initial selection of objects was limited by the morphology parameter $0.6 \leq \rho \leq 1.0$ \citep[see][]{2012AJ....143..123M}. From there, we chose objects with the difference between the minima depths smaller than $\delta_{min}=0.1$ in the units of the normalized flux. Next, we selected the objects flagged with the flat-bottom minimum (flag \textit{FB} in KEBC). This criterion is vital from the perspective of light curve numerical modeling. \citet{2003CoSka..33...38P} and then \citet{terrell} showed that the presence of a total eclipse in a contact binary allows to recover its mass ratio and inclination with a very high certainty. This allowed us to study the selected objects without the necessity of having their spectroscopic mass ratios.

\begin{figure*}
	\includegraphics[width=\linewidth]{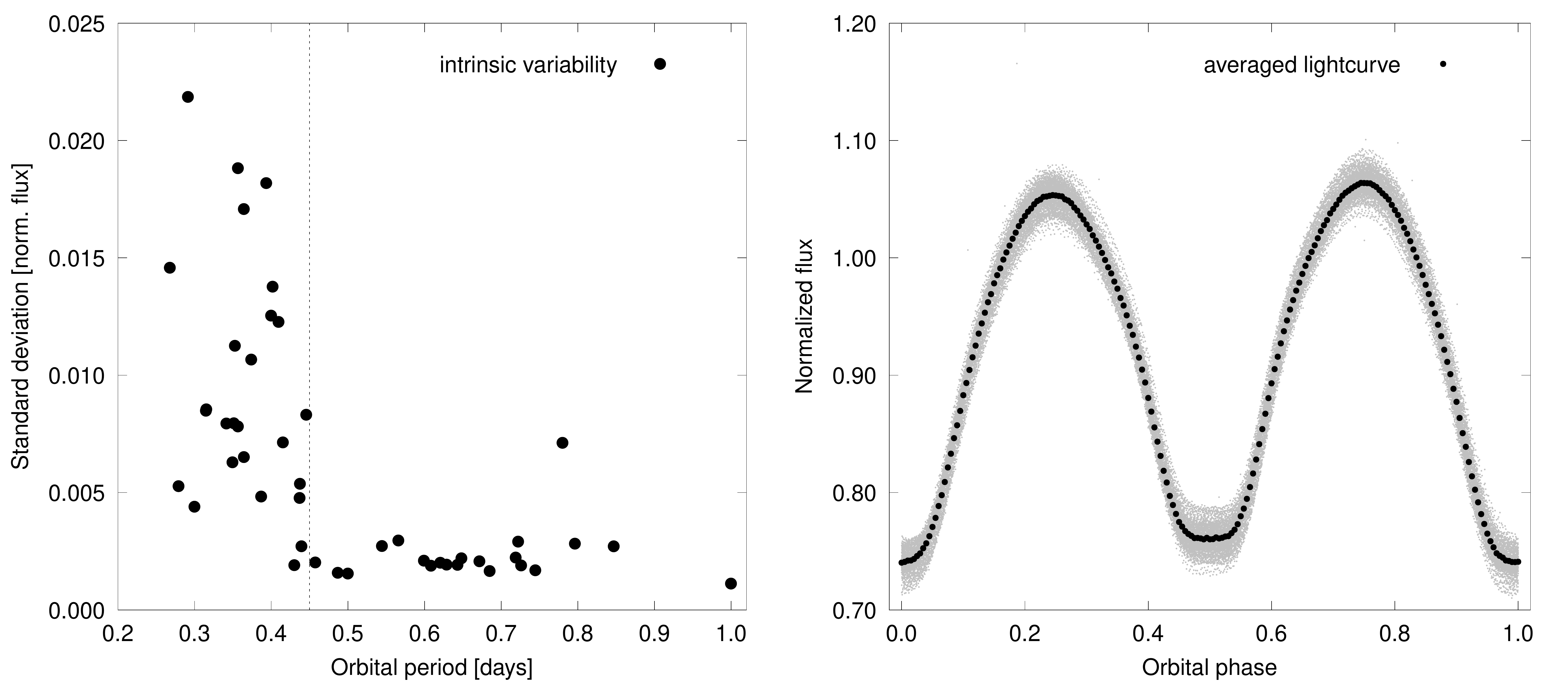}
    \caption{The intrinsic variability of light curves in our sample. The right panel shows an example of how the intrinsic variability produces a data scattering a phased light curve of KIC 2159783. The phased light curve (grey dots) is superimposed with its averaged counterpart (black dots). The left panes shows the distribution of the intrinsic variability along the orbital period for objects in our sample. The intrinsic variability was measured here as the standard deviation of the residuals left after subtracting the averaged light curve from the phased data.}
    \label{fig:intravar}
\end{figure*}

After downloading the light curves we conducted a visual inspection. For that we used the already detrended flux provided in KEBC. The visual inspection was possible with a prior phase-folding of each light curve. We adopted a bin width of approx. 12 orbital periods. The exact number or orbital periods used in this procedure varied at each object, as the light curves tend to experience different rates of intrinsic variability. Afterwards, because of the possible third light contamination, we gradually tightened the constraint on $\delta_{min}$. The extend of the constraint was determined individually per object, largely depending on the width of the flat-bottom minimum. This left us with a final set of 47 objects, of which study we present here. 

\section{Software}

The numerical modeling performed in this work utilized a modified Wilson-Devinney code. In our version of the code, the differential correction mechanism was replaced with a controlled Monte Carlo method. The same code was used earlier in, e.g., \citet{zola10,debski14,debski15,debski19,debski20}. For the numerical simulations we used a light curve generator based on the same version of the modified Wilson-Devinney code. This code is still using monochromatic wavelengths. Each and every contact binary model in this work assumes a square root limb darkening law with the limb darkening coefficients taken from \citet{claret2011,claret2013}. The gravitational brightening coefficient was fixed to $\beta=0.08$ (g$\,=0.32$) \citep[][]{lucy68a} and the albedo was fixed at $A=0.5$ \citep[][]{1969AcA....19..245R}

\begin{table}
    \centering
    \caption{Modeling results of the light curve distorted by an averaged longitudinal spot migration (SMA). The results are grouped by column, e.g. column one contains the results for the system with an input mass ratio $q=0.10$. The values are given in a relative form, i.e. as the ratio of the parameter value after the modeling to the input one. The apparent mass ratio and inclination are greater than their input counterparts, regardless the initial mass ratio. The apparent fill-out factor is significantly enlarged for the smallest input mass ratio only. The secondary component effective temperature is not affected by the SMA whatsoever.}
    \begin{tabular}{lccc}
\hline\hline
$q$ (model) & 0.10 & 0.20 & 0.30 \\
\hline
$\Delta i$   & 1.03 & 1.03 & 1.02 \\
$\Delta T_2$ & 1.00 & 1.00 & 1.00 \\
$\Delta ff$  & 1.11 & 1.01 & 1.00 \\
$\Delta q$   & 1.05 & 1.05 & 1.05 \\
    \end{tabular}

    \label{tab:sma}
\end{table}

\begin{table}
    \caption{Modeling results for the light curve distorted by the phase smearing effect (PS). The values are presented in the same relative fashion as in Table~\ref{tab:sma}. Results in this table are grouped by the input mass ratio. The columns represent subsequent runs with an increasing orbital period simulated in the light curve smearing. The modeling process did not take into account the phase smearing. Interestingly, the $\Delta q$ values are very similar in between the input mass ratios, unlike the $\Delta i$ and $\Delta ff$, which are affected the most for the lowest input mass ratios. PS does not affect the secondary component apparent effective temperature.}
    \begin{tabular}{llcccccc}
\hline\hline
&$P_{orb}$ [d]  & 0.25 & 0.30 & 0.35 & 0.40 & 0.60 & 0.80 \\
\hline
\multirow{4}{0.12\linewidth}{$q = 0.1$}&$\Delta i$  & 0.88 & 0.93 & 0.95 & 0.95 & 0.97 & 0.99 \\
&$\Delta T_2$&                                      1.00 & 1.00 & 1.00 & 1.00 & 1.00 & 1.00 \\
&$\Delta ff$ &                                      5.48 & 3.17 & 2.43 & 2.43 & 1.80 & 1.21 \\
&$\Delta q$  &                                      0.85 & 0.92 & 0.94 & 0.94 & 0.97 & 0.99 \\
\hline
\multirow{4}{0.12\linewidth}{$q = 0.2$}&$\Delta i$  & 0.92 & 0.95 & 0.96 & 0.96 & 0.99 & 0.99 \\
&$\Delta T_2$&                                      1.00 & 1.00 & 1.00 & 1.00 & 1.00 & 1.00 \\
&$\Delta ff$ &                                      3.23 & 2.26 & 2.16 & 2.16 & 1.29 & 1.07 \\
&$\Delta q$  &                                      0.88 & 0.93 & 0.95 & 0.95 & 0.98 & 0.99 \\
\hline
\multirow{4}{0.12\linewidth}{$q = 0.3$}&$\Delta i$  & 0.95 & 0.96 & 0.97 & 0.97 & 0.99 & 1.00 \\
&$\Delta T_2$&                                      1.00 & 1.00 & 1.00 & 1.00 & 1.00 & 1.00 \\
&$\Delta ff$ &                                      2.63 & 2.07 & 1.76 & 1.76 & 1.28 & 1.05 \\
&$\Delta q$  &                                      0.86 & 0.92 & 0.94 & 0.94 & 0.98 & 0.99 \\
    \end{tabular}
    \label{tab:ps}
\end{table}

\begin{table}
    \caption{Modeling results for the light curves distorted by both the phase smearing (PS) and the spot migration averaging (SMA) effects. The design of this table is exactly the same as Table~\ref{tab:ps}. The modeling did not account for the PS nor the SMA effects. For binaries with $P\geq0.3$, both PS and SMA are largely complementing each other in the mass ratio domain.}
    \begin{tabular}{llcccccc}
\hline\hline
&$P_{orb}$ [d]  & 0.25 & 0.30 & 0.35 & 0.40 & 0.60 & 0.80 \\
\hline
\multirow{4}{0.10\linewidth}{q = 0.1}&$\Delta i$  & 0.90 & 0.94 & 0.96 & 0.96 & 0.99 & 1.03 \\
&$\Delta T_2$&                                      1.00 & 1.00 & 1.00 & 1.00 & 1.00 & 1.00 \\
&$\Delta ff$ &                                      5.16 & 3.01 & 2.43 & 2.44 & 1.77 & 1.15 \\
&$\Delta q$  &                                      0.91 & 0.97 & 1.00 & 1.00 & 1.02 & 1.04 \\
\hline
\multirow{4}{0.10\linewidth}{q = 0.2}&$\Delta i$  & 0.94 & 0.96 & 0.98 & 0.98 & 1.00 & 1.03 \\
&$\Delta T_2$&                                      1.00 & 1.00 & 1.00 & 1.00 & 1.00 & 1.00 \\
&$\Delta ff$ &                                      3.08 & 2.48 & 1.97 & 1.97 & 1.39 & 1.04 \\
&$\Delta q$  &                                      0.93 & 0.97 & 1.00 & 1.00 & 1.03 & 1.04 \\
\hline
\multirow{4}{0.10\linewidth}{q = 0.3}&$\Delta i$  & 0.96 & 0.98 & 0.99 & 0.99 & 1.00 & 1.02 \\
&$\Delta T_2$&                                      1.00 & 1.00 & 1.00 & 1.00 & 1.00 & 1.00 \\
&$\Delta ff$ &                                      2.80 & 2.02 & 1.67 & 1.67 & 1.44 & 1.00 \\
&$\Delta q$  &                                      0.91 & 0.97 & 1.00 & 1.00 & 1.02 & 1.04 \\
    \end{tabular}
    \label{tab:smaps}
\end{table}

\section{The light curve intrinsic variability}
\label{sec:intrvar}

An average Kepler mission light curve consists of about 65\,000 data points, which, for a typical contact binary system, translates into about 4\,000 orbital epochs. A standard approach towards the numerical modeling of such a light curve require a prior phase folding and averaging. The result is a very convenient reduction of the number of data points. Only in such a way the numerical modeling codes \citep[e.g. the latest versions of the W-D code][]{2020ascl.soft04004W} can work within the reasonable time limits.
But phase folding over such many orbital epochs may cause the resulting phased light curve to be burdened with trace distortions that come from a short-time scale intrinsic variability. It is usually assumed that the process of averaging will safely remove any unwanted signal, as the intrinsic variability will cancel itself out over a long period of time. Or at least its impact will be minimized to the point of it being negligible. We decided to challenge this assumption. 

To start with, we measured the magnitude of the intrinsic variability occurring over four years of Kepler mission, for the objects in our sample. In the most basic way, the intrinsic variability can be expressed as a standard deviation of residuals left after subtracting the averaged light curve from the phase-folded data. A discussion on this method, as well as a presentation of an alternative approach, is available in the Appendix~\ref{appendix1}. We established that in our sample the intrinsic variability becomes an important factor in objects with orbital period $P<0.45$\,d. Interestingly, for objects with $P>0.45$\,d the intrinsic variability is almost negligible. As shown in the left panel of Fig.~\ref{fig:intravar}, the $P=0.45$\,d boundary divides the sample with a sharp cut-off. While this boundary is interesting by itself, we decided to tackle its physical implications in an another study.

A proper analysis of the Kepler light curve distortions has to acknowledge the 0.5 hour integration time of the Kepler Long Cadence mode, which introduces the `phase smearing` effect. Its impact on a light curve can partially be taken into account when using the newest versions of the original W-D code. \citet{zola2017} showed that for contact binaries with orbital period shorter than $P=0.4$\,d it is absolutely imperative to take into account the phase smearing effect during the numerical modeling of the Kepler Mission light curves. We will revisit that analysis in this Section, as well as confront the phase smearing with the averaging of the intrinsic variability.

To quantify the impact of the intrinsic variability averaging on the numerical modeling results, we devised a set of modeling runs based on three synthetic contact binaries. The simulated systems were differing in mass ratios: $q=0.10,\,0.20,\,0.30$ and had otherwise identical system parameters: $i=85^{\circ}$, $T_1=T_2=6\,000$\,K, $ff=0.1$. The fill-out factor $ff$ is defined here as:

$$ff = \frac{\Omega_{L_1}-\Omega}{\Omega_{L_1}-\Omega_{L_2}},$$ 

\noindent where $\Omega_{L_1}$ is the Roche pseudopotential at the inner critical Lagrange surface, $\Omega_{L_2}$ - analogously at the outer critical Lagrange surface, and $\Omega$ is the pseudopotential of the system (i.e. the surface of the binary). 

\subsection{Averaged spot migration}
\label{sec:sma}

We assumed the intrinsic variability in contact binaries is caused entirely by the starspot migration. We adopted a high-latitude, large spot, following our preliminary analysis \citep[][]{debski14,2014bsee.confP..42D,debski20} as well as the findings in \citet{tran2013}. We recreated the light curve intrinsic variability by simulating longitudinally moving spot with a fixed size, temperature, and latitude. Spot position differed by $10^{\circ}$ in subsequent light curves. After obtaining a full $360^{\circ}$ longitudinal migration, we averaged all 36 light curves into one, Spot Migration-Averaged (SMA) light curve. We followed this procedure for all three base systems ($q=0.1,\,0.2,$\,and $q=0.3$).

Next, we performed a numerical modeling on the SMA light curves. The modeling was designed to not compensate for the intrinsic variability averaging. The results are presented in Table~\ref{tab:sma}. The quantities in the table are relative, and should be read as a ratio of the modeled parameter value to the base (input) value, such as:

$$ \Delta x \equiv \frac{x_{result}}{x_{input}},$$

\noindent where ${x_{input}}$ is a value of parameter $x$ used when simulating the base system, and ${x_{result}}$ is a value of the same parameter, obtained after the numerical modeling of the SMA light curve of a said system. Such presentation will ease the comparison of the results between cases with different distorting agents present in simulated light curves. The formal errors of the modeled parameters are no higher than 1\%.

The simulated averaged intrinsic variability has increased the apparent (obtained from modeling) inclination and mass ratio in all three cases (see Table~\ref{tab:sma}). Interestingly, the increase is (almost) uniform, disregarding the mass ratio of the base system. The apparent fill-out factor has risen only for the system with the lowest mass ratio. The temperature of the secondary component remained unchanged.

\subsection{Phase smearing revisited}
\label{sec:ps}

The next run concerned the influence of the phase smearing (PS). This part is in essence an extension of the original \citet{zola2017} study, performed on our three base systems. We introduced the phase smearing to the light curves, as if the systems were observed with the Kepler Long Cadence mode (0.5 h integration time). Same kind of phase smearing occurs also in the TESS mission light curves. The phase smearing was simulated with the assumption of six different orbital periods: $P=0.25$, 0.30, 0.35, 0.4, 0.6, and $P=0.8$\,d. In this run we did not simulated the intrinsic variability. During the modeling process, we did not compensate for the phase smearing.

The outcome of the numerical modeling of the phase-smeared light curves are showed in Table~\ref{tab:ps}. The results are presented, again, in relative values. In contrast to Table~\ref{tab:sma}, the results in Table~\ref{tab:ps} are grouped by rows, according to the input mass ratio of the system. Columns represent different magnitudes of the phase smearing i.e. the subsequently increasing orbital periods. For example, the phase smearing causes the numerically modeled apparent inclination in the $q=0.1$ system with orbital period adopted as $P=0.25$\,d, to be 0.88 of the original inclination in the base system (i.e. 12\% lower). Under the same circumstances, the apparent mass ratio is only 0.85 of the input (true) mass ratio of the system.

As expected, the phase smearing affects the parameters the most for the shortest orbital periods. The disruption of the light curve causes a chain reaction of the modeled system parameters. Smoothing of the light curve around the secondary minimum forces the inclination to drop. This is followed by a compensation of the light curve amplitude, modified by the fill-out factor. The highly thinned-out light curve maxima are being fitted with lowering the mass ratio. Finally, because that would produce more noticeable first and fourth eclipse moments, the fill-out factor plays an another role of smoothing the light curve in between the maxima and minima. Interestingly, because at no point the light curve minima change their depths with respect to each other, the temperature of the secondary component remains unchanged.

\subsection{Averaged spot migration and phase smearing combined}
\label{sec:smaps}

In the final run we recreated the light curve distortion caused by the averaged intrinsic variability and phase smoothing simultaneously. The light curve preparation was similar to the one in Section~\ref{sec:ps}, but this time the spot migration was introduced and its effect is averaged in the same fashion as in Section~\ref{sec:sma}. The results are stored in Table~\ref{tab:smaps}. Just as in the last two runs, all the best fitting models we obtained, were of contact configuration ($ff>0$).

The cumulative impact of both effects produces a very interesting pattern. Apart from the extremely short orbital periods, the apparent mass ratio is barely affected. This happens regardless the true mass ratio of the base system. The difference between the apparent and true mass ratio for contact binaries with orbital period of $P=0.3$\,d is just 3\%. Such a difference is on the edge of the usual precision of numerical modeling. On the other hand, the decrease of the apparent mass ratio still has to be taken into account for binaries with $P<0.3$\,d.

In the simulated objects with the orbital periods $P>0.6$\,d the apparent mass ratio outgrew the input mass ratio. But as we established earlier, in the real world-scenario the intrinsic variability becomes unimportant right above $P=0.45$\,d. That means that the superposition of the PS and SMA effects will occur only below that threshold, leaving the $P>0.45$\,d objects under the influence of the PS only. This is rather fortunate, since the PS looses its impact on the light curve quickly with a rising orbital period. The distortion of the observed mass ratio is therefore naturally diminished and remains non-negligible only fot the extremely short-period contact binaries, close to $P = 0.25$\,d.

The only one remaining highly affected system parameter is the fill-out factor. In the extreme cases of the lowest mass ratio and shortest orbital period, its apparent value was as much as five times larger than the input. One should remember that input fill-out factor in our simulated light curves was just $ff=0.1$. In the future, a follow-up investigation should be performed for a higher value input fill-out factor.

\begin{table*}
	\caption{Modeling result for objects coinciding with the \texttt{Zola17} sample. The results from this work are flagged as row '1', while the archival results are in the row '2'. The values from this work come from the best fitting model from the MC search, hence have no formal errors. The errors for the \texttt{Zola17} are presented in brackets and apply to the last significant digits. The accounting for the phase smearing introduced in \texttt{Zola17} work produces almost the same result as in our work. The distinction between two approaches is more visible in the necessity of the third light introduction. However, if the third light had to be introduced in both approaches, its values are mostly similar. The temperature ratios between the two approaches are extremely similar, as predicted in Table~\ref{tab:smaps}. Interestingly, the fill-out factor as well do not differ in between the two approaches. A visualization of selected relations between modeled parameters is presented in Fig.~\ref{fig:47corel}.}
	\label{tab:results_comparison}
	\begin{tabular}{ccclclclllll}
	\hline
    KIC \# & $P_{orb}$ [d] && $i$ [$^{\circ}$] & $T_1$ [K] & $T_2$ [K] & $T_2/T_1$ & $\Omega_{1,2}$ & $ff$ [\%] & $q$ & $L_1$/($L_1+L_2$) & $l_3$ \\
    \hline
\multirow{2}{*}{3104113} & \multirow{2}{*}{0.8467860} & 1 & 80.0(8) &  6535 &   6619(28) & 1.013 & 2.0644(162) &  93(14) & 0.1748(49) & 0.7992(1)  & 0.0 		 \\\vspace{.2cm}
						 & & 2 & 79.05(11) & 5910 & 5994(1) & 1.014 & 2.0498(4) & 91(1) & 0.1666(2) & 0.8027(5) & 0.0 	  \\
\multirow{2}{*}{3127873} & \multirow{2}{*}{0.6715256} & 1 & 87.6(1.0) &  6408 &   6168(34) & 0.963 & 1.9171(35) &  99(5) & 0.1094(30) & 0.8767(22)  & 0.284(25)       \\\vspace{.2cm}
                         & & 2 & 90.00b	 & 6070 & 5702(3) & 0.939 & 1.9242(2) & 88(1) & 0.1093(9) & 0.8837(31)& 0.215(2) \\                                                                                        
\multirow{2}{*}{5439790} & \multirow{2}{*}{0.7960862} & 1 & 83.0(9) &  7022 &   6804(23) & 0.969 & 2.1765(95) &  39(8) & 0.1969(28) & 0.8208(1)  & 0.0          \\\vspace{.2cm}
						 & & 2 & 82.69(2)  & 6566 & 6411(1) & 0.976 & 2.1686(2) & 36(1) & 0.1921(1) & 0.8237(6) & 0.0	  \\ 				 				   					   				   			   		
\multirow{2}{*}{5809868} & \multirow{2}{*}{0.4393902} & 1 & 78.7(5) &  7208 &   6484(23) & 0.900 & 2.0540(54) &  37(6) & 0.1446(19) & 0.8850(1)  & 0.0         \\\vspace{.2cm}
						 & & 2 & 90.00b    & 6880 & 6365(1) & 0.925 & 2.1743(4) & 47(1) & 0.2007(2) & 0.8454(11)& 0.208(1)	\\ 				 				   					   				   			   		
\multirow{2}{*}{7698650} & \multirow{2}{*}{0.5991551} & 1 & 81.8(1.8) &  6307 &   6261(23) & 0.993 & 1.9516(145) &  74(19) & 0.1161(57) & 0.8638(26)  & 0.133(27)       \\\vspace{.2cm}
						 & & 2 & 85.35(8)  & 6110 & 6082(1) & 0.995 & 1.9720(10)  & 70(1) & 0.1232(3) & 0.8576(22)& 0.159(2)	\\				 				   					   				   			   		
\multirow{2}{*}{8145477} & \multirow{2}{*}{0.5657843} & 1 & 84.3(1.8) &  6538 &   6284(43) & 0.961 & 1.9068(136) &  75(21) & 0.0988(50) & 0.8912(31)  & 0.180(30)       \\\vspace{.2cm}
						 & & 2 & 90.00b 	 & 6800 & 6496(2) & 0.955 & 1.9220(10)  & 65(1) & 0.1020(3) & 0.8933(33)& 0.159(2)	\\   				 				   					   				   			   		
\multirow{2}{*}{8265951} & \multirow{2}{*}{0.7799575} & 1 & 79.7(6) &  6943 &   6648(16) & 0.958 & 2.0759(64) &  40(6) & 0.1546(18) & 0.8542(1)  & 0.0       \\\vspace{.2cm}
						 & & 2 & 79.38(4)  & 7044 & 6780(1) & 0.963 & 2.0760(10)  & 38(1) & 0.1540(1) & 0.8565(2) & 0.0		\\ 				 				   					   				   			   		
\multirow{2}{*}{8539720} & \multirow{2}{*}{0.7444991} & 1 & 82.4(1.9) &  6658 &   6402(43) & 0.962 & 2.0114(243) &  90(25) & 0.1480(110) & 0.8469(70)  & 0.475(28)       \\\vspace{.2cm}
						 & & 2 & 85.11(4)  & 6350 & 6119(1) & 0.964 & 2.0378(3) & 86(1) & 0.1581(2) & 0.8426(19)& 0.484(1)	\\	

\multirow{2}{*}{8804824} & \multirow{2}{*}{0.4574038} & 1 & 88.6(1.7) &  6556 &   6187(39) & 0.944 & 1.9386(134) &  75(19) & 0.1111(48) & 0.8884(35)  & 0.240(29)        \\\vspace{.2cm}
						 & & 2 & 90.00b 	 & 7200 & 6733(2) & 0.935 & 1.9438(6) & 67(1) & 0.1109(3) & 0.8937(26)& 0.192(2) \\ 				 		

\multirow{2}{*}{9350889} & \multirow{2}{*}{0.725948} & 1 & 87.4(1.1) &  6996 &   7037(41) & 1.006 & 2.0129(74) &  90(8) & 0.1485(29) & 0.8248(22)  & 0.237(16)       \\\vspace{.2cm}
						 & & 2 & 79.92(2)  & 6725 & 6749(2) & 1.004 & 1.9173(2) & 87(1) & 0.1060(1) & 0.8702(9) & 0.076(1)	\\ 				 		
\multirow{2}{*}{9453192} & \multirow{2}{*}{0.7188371} & 1 & 85.1(1.9) &  6622 &   6161(33) & 0.930 & 2.0196(169) &  61(19) & 0.1395(75) & 0.8761(37)  & 0.241(31)        \\\vspace{.2cm}
						 & & 2 & 89.51(6)  & 6730 & 6239(1) & 0.927 & 2.0540(10)  & 62(1) & 0.155(1)  & 0.8793(14)& 0.269(1)	\\			

\multirow{2}{*}{10007533} & \multirow{2}{*}{0.6480635} & 1 & 81.7(1.6) &  6977 &   6388(31) & 0.916 & 1.8694(94) &  54(18) & 0.0808(19) & 0.9210(1)  & 0.0  \\\vspace{.2cm}
                          & & 2 & 90.00b 	 & 6810 & 6356(1) & 0.933 & 1.9126(3) & 76(1) & 0.1011(1) & 0.9013(19)& 0.178(1) \\             

\multirow{2}{*}{10229723} & \multirow{2}{*}{0.6287243} & 1 & 81.5(1.7) &  6477 &   6262(28) & 0.967 & 2.0408(185) &  43(20) & 0.1418(77) & 0.8616(27)  & 0.168(29)        \\\vspace{.2cm}
                          & & 2 & 83.16(5) & 6201 & 6000(1) & 0.968 & 2.0570(10)  & 36(1) & 0.145(1)  & 0.8614(18)& 0.166(1)	\\              

\multirow{2}{*}{10267044} & \multirow{2}{*}{0.4300365} & 1 & 79.4(5) &  7103 &   6900(24) & 0.971 & 2.1337(79) &  55(7) & 0.1865(26) & 0.8229(1)  & 0.0          \\\vspace{.2cm}
                          & & 2 & 89.56(7) & 6808 & 6700(1) & 0.984 & 2.2463(1) & 55(1) & 0.240(1)  & 0.7828(9) & 0.150(1)	\\            

\multirow{2}{*}{11097678} & \multirow{2}{*}{0.9997156} & 1 & 83.5(2.0) &  6493 &   6425(48) & 0.990 & 1.8878(145) &  90(23) & 0.0951(51) & 0.8804(41)  & 0.268(31)        \\\vspace{.2cm}
                          & & 2 & 85.14(2) & 6493 & 6426(1) & 0.990 & 1.8928(1) & 87(1) & 0.0967(1) & 0.8792(5) & 0.267(1)	\\       

\multirow{2}{*}{11144556} & \multirow{2}{*}{0.6429797} & 1 & 76.8(1.3) &  6803 &   6688(46) & 0.983 & 2.0304(180) &  98(17) & 0.1608(92) & 0.8229(35)  & 0.380(21)         \\\vspace{.2cm}
                          & & 2 & 76.84(2) & 6428 & 6318(1) & 0.983 & 2.0424(2) & 97(1) & 0.1607(1) & 0.8246(6) & 0.370(1)	\\        

\multirow{2}{*}{12055014} & \multirow{2}{*}{0.4999046} & 1 & 85.3(1.5) &  6448 &   6356(30) & 0.986 & 2.0466(134) &  73(13) & 0.1562(62) & 0.8345(25)  & 0.132(27)       \\\vspace{.2cm}
						  & & 2 & 90.00b 	 & 6456 & 6439(1) & 0.997 & 2.0606(1) & 67(1) & 0.1598(1) & 0.8346(9) & 0.120(1)	\\
\hline
	\end{tabular}
\end{table*}
\section{The numerical modeling}
\label{sec:modeling}

\subsection{Numerical modeling scheme}
To recapitulate, the primary goal of this study was to establish whether the objects in our sample are contact binaries, and to establish their photometric mass ratios. We began with rephasing the light curves so that the total eclipse minimum would correspond to $\phi=0.5$. With that we could perform a numerically uniform search with a mass ratio $q=\frac{M_2}{M_1}<1.0$. We averaged the phased curves with a $0.005\,\phi$ bin size. From now on we will refer to such averaged phased light curves simply as to the 'light curves'. In the next step we divided the sample into two subsets, depending on the presence of the O'Connell effect. The light curves with unequal brightness maxima heights, were modeled with a dark, circular starspot residing on the primary (i.e. more massive) companion. The searched spot parameters were: spot center longitude $\theta_s$ and co-latitude $\lambda_s$, and spot radius $r_s$. Spot relative temperature was fixed at 0.75 of the surrounding effective temperature, $T_1$. The choice of a cool, dark spot was dictated by its presumed magnetic origin. The objects without the visible O'Connell effect were subdued to the spot-free light curve modeling. Each model had a fixed temperature of the primary component. We took the temperatures from the KEBC, as it was the most complete and uniform source at the time. If the modeling process has ended without reaching the experimentally adopted threshold or if it did not achieve convergence at all, we repeated the modeling with a third light added to the search.
Our program did not account for the phase smearing, nor for the averaged intrinsic variability effects. The two main reasons for such course were:

\noindent 1)~the two effects are largely complementing each other in affecting the mass ratio (only four of our objects have orbital period $P~\leq~0.3$~d),

\noindent 2)~it was interesting to compare our results with \citet{zola2017}, who took the phase smearing into account during the modeling process in their sample of contact binaries (hereafter referred to as \texttt{Zola17} sample).

All 47 models best fitting to our light curves returned a contact configuration. 21 light curves exhibited the O'Connell effect, and therefore were modeled with a dark spot. Out of remaining 26 objects, 17 are consisting the \texttt{Zola17} sample. Here we are dividing our sample into three sets, which are presented below.

\subsection{Modeling results of objects with the O'Connell effect}

We modeled 21 objects which needed an introduction of a starspot. 12 of them required additional third light to fit the model to data accurately. The results of the modeling are stored in Table~\ref{tab:results_spot}. For objects with an orbital period $P<0.3\,$d we calculated a `corrected mass ratio`, $q_{corr}$. This was an attempt to utilize our findings on the combined intrinsic variability averaging and phase smearing effects. $q_{corr}$ was calculated as a two dimensional ($q,\,P$) linear interpolation based on Table~\ref{tab:smaps}.

\subsubsection{Preferred spot longitude}
\label{sec:model-oconnell}

In 19 out of the 21 objects the modeled cool spot resides on the trailing side of the binary. The mean longitude for such a spot is $\lambda_s=302^{\circ} \pm 36^{\circ}$). The two remaining binaries, KIC 11618883 and KIC\,8554005 have spots modeled at longitudes $\lambda_s=94.4^{\circ}$ and $\lambda_s=119.7^{\circ}$, respectively. Both of them have orbital periods far greater than $P=0.45\,$d cutoff ({$P=0.6848719$\,d} and {$P=0.6083407$\,d} respectively). The one system with a spot on a trailing side and the orbital period exceeding the cutoff is KIC 5290305 ($P=0.6205088$\,d). Its spot is modeled at the longitude $\lambda_s=211.4^{\circ}$, which is the third lowest longitude in the set, right after the two already mentioned systems. 

\begin{figure*}
    \centering
    \includegraphics[width=0.85\linewidth]{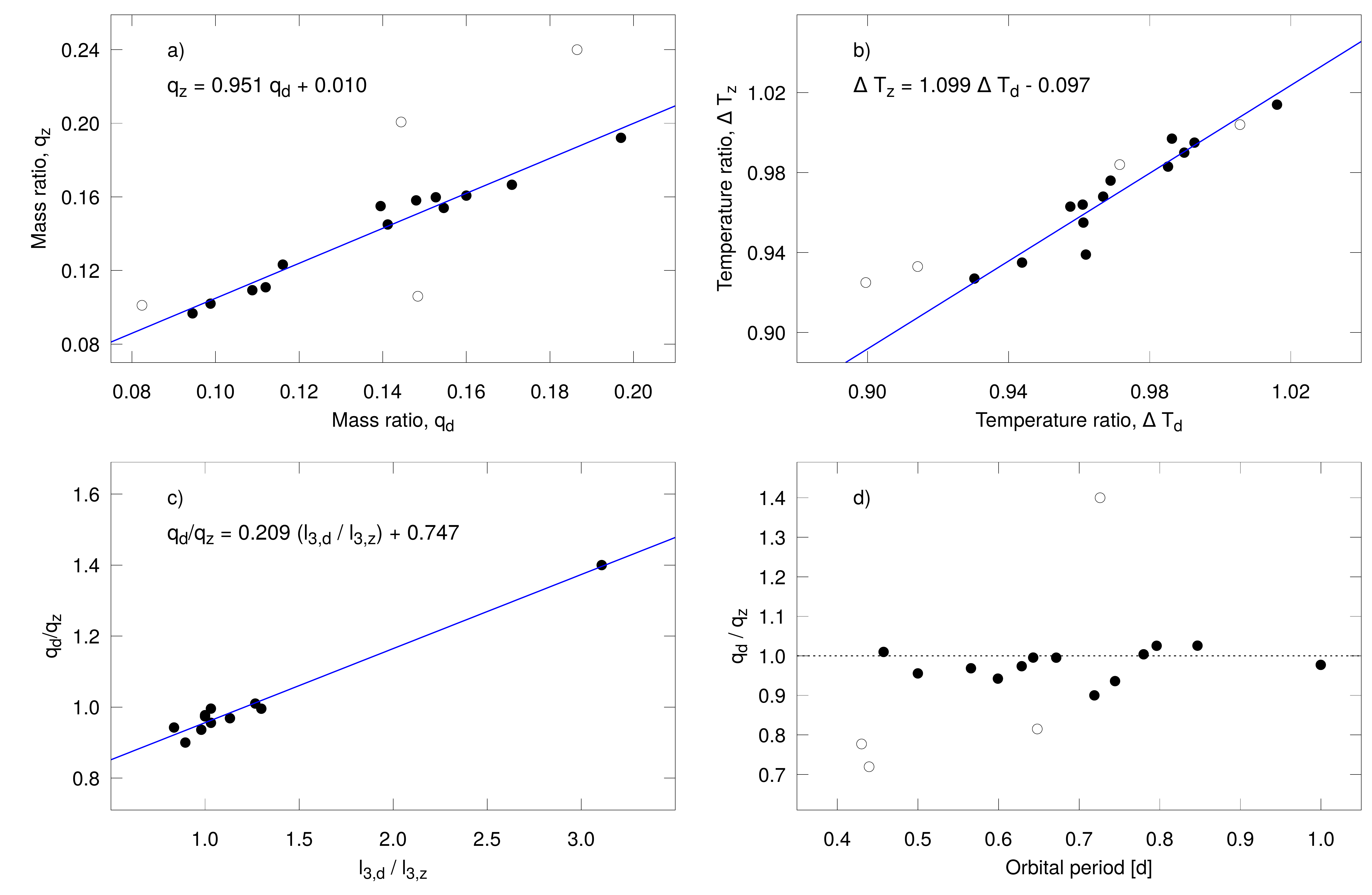}
    \caption{Relation between the modeled parameters obtained independently by \texttt{Zola17} and in this work. Panel 'a' shown a correlation between the mass ratios and a linear fit that parametrizes the relation. The fit was made disregarding the objects for which the third light was significantly different between the methods (see the discussion). These are marked with an empty circle. Panel 'b' follows the same construct as the previous one, but depicts a correlation between the temperature ratios. Panel 'c' focuses on the relation between the ratio of mass ratios between the methods and ratio of the third light, also between the methods. Note the outlying point comes from KIC\,9350889 and in this panel is not disregarded from the analysis. However, the objects for which the third light was zero in at least one method, are discarded. Panel 'd' shows the distribution of the ratio of mass ration between methods, along the orbital period. The dotted line marks the ratio equal to one.}
    \label{fig:badezola}
\end{figure*}

\subsection{Reexamination of the \texttt{Zola17} sample}

The second set consists of 17 binaries studied earlier with a phase smearing taken into account. Almost all of them fall above the $P=0.45\,$d limit, and are therefore bound to have negligible intrinsic variability averaging effect. The remaining two objects are just on the inner edge of the $P=0.45\,$d limit: KIC 5809868 ($P=0.4393902$\,d) and KIC 10267044 ($P=0.4300365$\,d). These objects are also the only two with such a large difference in the relative minima depth.

The results of our modeling are superimposed with the \texttt{Zola17} results in Table~\ref{tab:results_comparison} and grouped per object. In addition to the effective temperatures, we are showing the temperatures ratio: 
$$\Delta T = \frac{T_2}{T_1}$$
Without such an approach it would be impossible to compare our results with the archival one, since the latter used different effective temperatures for the primary component.

Most of the mass ratios obtained in this work are in agreement with those from \texttt{Zola17}. The objects in which $q$'s are in high disagreement are those in which there is a large discrepancy in the third light. There are four of such objects. In three of them \texttt{Zola17} added a third light into a model, while we did not see that as a necessity. Two of them are the already mentioned KIC 5809868 and KIC 10267044. The third one is KIC\,10007533, which has the lowest mass ratio in the whole set. The last object with a large third light difference is KIC\,9350889. The only existing explanation for the discrepancy between the best models for this object's light curve is that neither our, nor the \texttt{Zola17} models fit the descending arm of the secondary minimum very well. This overshooting of the models might point toward a yet poorly understood fine light curve asymmetry worth investigating.

Not counting the four above objects, the mass ratios established with our method are highly similar to those obtained by \texttt{Zola17}. Exactly the same can be said about the temperature ratios and, surprisingly, about the fill-out factors. A depiction of the correlations between the parameters obtained with two methods presented in Fig~\ref{fig:47corel}. Taking into account the phase smearing during a modeling process had no influence over the best fitting models for the majority of objects in \texttt{Zola17} sample. It might be important for objects with a significantly shorter orbital period, such as $P \approx 0.25$\,d.

However, compensating solely for the phase smearing without tackling the intrinsic variability for objects below the $P=0.45$\,d limit might result in a faulty model. In \texttt{Zola17} sample both such objects were modeled with an additional third light, which turned out to be unnecessary in our study. Moreover, we do not see any light time effect in the minima timing diagrams on either of the objects. There are also no other stars significantly close to these objects, even considering the rather large Kepler Mission pixel size. We could conclude that accounting solely for the phase smearing in case of $P<0.45$\,d binaries can be harmful to the model, but at this point we are working on just two objects. This issue is certainly interesting as it might additionally prevent an overestimation of the multiplicity of contact binaries companions.

\subsection{Modeling results of the nine systems with no prevailing spot}

The last set consisted of nine systems with no O'Connell effect in the averaged light curve. These were not included in \citet{zola2017}, hence are presented separately. Three of them are above the $P=0.45\,$d cutoff and therefore experience next to none intrinsic variability. Remaining six objects show moderate and heavy intrinsic variability which largely averages itself out. The resulting best fit model parameters are given in Table~\ref{tab:results_nospot}. Five out of nine objects needed an introduction of a third light.

\begin{table*}
	\caption{Modeling results for objects with a visible O'Connell effect. The light curves of objects in this set were modeled with a cool spot on a surface of a primary component. The temperature of a spot was fixed to $T_s=0.75\,T_1$ and therefore is not included in this table. The primary component effective temperature was the only fixed parameter, per the normal use of W-D code. The $q_{corr}$ are the modeled mass ratios of the objects with $P<0.3$\,d.}
	\label{tab:results_spot}
	\begin{tabular}{p{1.cm}p{1.1cm}p{0.8cm}clllp{1.1cm}p{1.2cm}ccccc}
	\hline
	    KIC \#  & $P_{orb}$ [d] & $i$ [$^{\circ}$] & $T_1$ [K] & $T_2$ [K] & $\Omega_{1,2}$ & $ff$ & $q$ & $L_1$/($L_1$+$L_2$) & $l_3$ & $\theta_{s}$[$^{\circ}$] & $\lambda_{s}$[$^{\circ}$] & $r_{s}$[$^{\circ}$]  \\
\hline
  2159783 &  0.3738842 & 80.0(6) &  6140 &   6339(18)    & 2.027(7)  &  68(8)  & 0.1455(2)  & 0.82(6)   & 0.0      &  20(3)  & 336(4)  & 17(2)  \\
  2437038 &  0.2676785 & 82.3(1.7) &  5461 &   5973(64)  & 2.066(20) &  86(18) & 0.1715(9)  & 0.75(9)   & 0.44(2)  &  59(16) & 295(10) & 13(4)  \\
  2570289 &  0.2790278 & 86.4(2.1) &  6360 &   6487(205) & 1.935(21) &  91(29) & 0.1143(9)  & 0.85(21)  & 0.70(2)  &  72(24) & 333(45) & 10(8)  \\
  3342425 &  0.3934148 & 87.7(9) &  6306 &   6462(38)    & 1.906(5)  & 100(7)  & 0.1050(2)  & 0.85(7)   & 0.10(2)  &  41(11) & 293(15) & 11(6)  \\
  4036687 &  0.2997993 & 82.4(1.4) &  6200 &   6381(52)  & 1.981(13) &  96(14) & 0.1367(6)  & 0.82(8)   & 0.27(3)  &  15(13) & 330(30) & 17(9)  \\
  4244929 &  0.3414027 & 85.9(1.4) &  5976 &   6053(122) & 1.995(15) & 100(16) & 0.1447(8)  & 0.82(13)  & 0.53(1)  &   9(16) & 233(48) & 22(7)  \\
  5283839 &  0.3152311 & 86.5(1.3) &  6239 &   6541(89)  & 1.955(9)  &  91(11) & 0.1229(4)  & 0.83(8)   & 0.22(2)  &  30(16) & 303(29) & 14(6)  \\
  5290305 &  0.6205088 & 81.6(1.3) &  6542 &   6188(45)  & 2.202(19) &  43(14) & 0.2106(10) & 0.83(7)   & 0.18(2)  &  33(14) & 211(6)  & 14(4)  \\
  6118779 &  0.3642464 & 78.4(6) &  5715 &   6003(26)    & 1.913(8)  &  93(11) & 0.1060(2)  & 0.84(6)   & 0.0      &  27(8)  & 342(4)  & 14(3)  \\
  7821450 &  0.3147619 & 81.4(1.5) &  5155 &   5513(37)  & 2.174(15) &  38(12) & 0.1955(7)  & 0.76(8)   & 0.32(1)  &  29(14) & 295(25) & 10(7)  \\
  8143757 &  0.3565228 & 79.9(5) &  5454 &   5592(23)    & 2.287(10) &  23(7)  & 0.2368(3)  & 0.76(6)   & 0.0      &   9(8)  & 266(19) & 15(5)  \\
  8432859 &  0.3511705 & 84.8(1.2) &  6352 &   6537(54)  & 1.957(8)  &  92(10) & 0.1241(4)  & 0.83(8)   & 0.20(2)  &  19(13) & 324(28) & 16(8)  \\
  8554005 &  0.6083407 & 83.5(1.4) &  7298 &   7303(18)  & 2.457(33) &  59(15) & 0.3562(22) & 0.70(7)   & 0.18(3)  &  39(11) & 120(3)  & 10(4)  \\
  8682849 &  0.3525548 & 78.9(7) &  5631 &   5831(14)    & 1.958(6)  &  80(8)  & 0.1206(2)  & 0.84(6)   & 0.0      &  22(6)  & 346(4)  & 16(3)  \\
  8842170 &  0.3493917 & 79.2(1.4) &  5589 &   5836(47)  & 1.981(17) &  96(19) & 0.1365(7)  & 0.81(10)  & 0.39(2)  &  11(16) & 295(13) & 22(6)  \\
  9087918 &  0.4456069 & 80.0(4) &  6085 &   6161(27)    & 2.305(9)  &  32(6)  & 0.2512(4)  & 0.76(6)   & 0.0      &   6(7)  & 302(15) & 22(5)  \\
  9283826 &  0.3565238 & 80.0(7) &  5987 &   6500(35)    & 2.072(10) &  70(9)  & 0.1663(3)  & 0.77(6)   & 0.0      &   9(10) & 301(9)  & 23(4)  \\
 10322582 &  0.2912692 & 75.2(5) &  5863 &   6537(28)    & 1.966(10) &  83(12) & 0.1250(3)  & 0.79(6)   & 0.0      &  27(6)  & 330(3)  & 20(2)  \\
 10528299 &  0.3998018 & 80.9(5) &  6302 &   6631(41)    & 2.178(10) &  42(7)  & 0.1998(3)  & 0.77(6)   & 0.0      &  10(11) & 272(13) & 21(4)  \\
 10618253 &  0.4374028 & 84.3(6) &  6580 &   6574(16)    & 1.945(4)  &  84(6)  & 0.1165(1)  & 0.86(7)   & 0.0      &  16(3)  & 340(6)  & 16(3)  \\
 11618883 &  0.6848719 & 87.6(1.6) &  4347 &   4403(34)  & 2.181(23) &  80(16) & 0.2253(11) & 0.76(14)  & 0.58(1)  &  10(13) &  94(24) & 26(5)  \\

\hline
	\end{tabular}
\end{table*}

\begin{table*}
	\caption{The results of a light curve modeling for objects in which the averaged phase curve did not showed any O'Connell effect. The primary component temperature was a fixed parameter during the modeling process.}
	\label{tab:results_nospot}
	\begin{tabular}{ccccccccccc}
	\hline
    KIC \# & $P_{orb}$ [d] & i [$^{\circ}$] & $T_1$ [K] & $T_2$ [K] & $\Omega_{1,2}$ & $ff$ & $q$ & $L_1$/($L_1$+$L_2$) & $l_3$  \\
\hline
  7601767 &  0.4867339 & 77.4(9) 	&  6567 & 6388(16) & 2.108(9)  &  23(9)  & 0.161(4) & 0.8483(11) & 0.25(2)  \\
  7709086 &  0.4094746 & 75.7(4) 	&  6108 & 6290(14) & 2.159(7)  &  37(6)  & 0.189(3) & 0.7930(2)  & 0.0 		\\
  8265951 &  0.7799575 & 79.7(6) 	&  6943 & 6648(16) & 2.076(6)  &  40(6)  & 0.155(2) & 0.8542(1)  & 0.0 		\\
  8496820 &  0.4369669 & 82.5(8) 	&  6601 & 6558(21) & 2.080(10) &  76(9)  & 0.173(3) & 0.8169(1)  & 0.0 		\\
  9030509 &  0.4017846 & 87.2(7) 	&  5326 & 5393(14) & 2.058(10) & 100(9)  & 0.176(5) & 0.7927(14) & 0.30(1)  \\
  9151972 &  0.3867961 & 87.7(1.5)  &  5622 & 5716(36) & 2.041(15) & 100(14) & 0.167(8) & 0.7982(55) & 0.54(2)  \\
  9703626 &  0.4151477 & 76.6(4) 	&  6060 & 6088(16) & 2.214(10) &  36(7)  & 0.212(4) & 0.7927(2)  & 0.0 		\\
  9776718 &  0.5443510 & 81.3(6) 	&  7205 & 7277(18) & 2.033(5)  &  86(5)  & 0.156(2) & 0.8177(1)  & 0.0 		\\
 10395609 &  0.3642540 & 87.2(1.4)  &  6566 & 6564(34) & 2.043(12) &  99(12) & 0.168(7) & 0.8096(25) & 0.21(3)  \\
 12352712 &  0.7220650 & 89.4(1.4)  &  6667 & 6469(35) & 1.890(8)  &  87(13) & 0.095(3) & 0.8877(22) & 0.21(2)  \\
\hline
	\end{tabular}
\end{table*}

\subsection{Contact binaries with a companion}

28 out of 47 contact binaries in this study needed a third light introduced to the model. Four of them are most probably due to the pixel light contamination. Two of such, KIC\,2437038 and KIC\,2570289 are located in a dense field of NGC 6791. The next two, KIC\,11097678 and KIC\,11618883 have considerably bright companions, which could not be resolved with the large Kepler Mission pixel size. None of the four binaries show any light-time effect in their minima timing.

In contrast, we found that KIC\,8265951 has a considerable light-time effect in the minima timing, but its best fitting model shows no third light. Two other objects: KIC\,10528299 and KIC\,7709086 are also modeled with no third light, but their minima timing diagrams show a possible light time effect.

There are additionally three binaries with some possibility of the pixel contamination by a nearby source: KIC\,3127873, KIC\,8432859, and KIC\,8842170. These objects have also no visible light-time effect. Assuming they are indeed contaminated, the lower limit for a fraction of binaries with a possible companion is 53\%. If we count in the three uncertain objects, then the fraction rises to 60\%. These limits fall close to the estimates made by \citet{2006AJ....131.2986P}. In their work the estimated fraction of contact binaries with a companion was about 59\% with a lower limit of about 40\%.

\begin{figure*}
    \centering
    \includegraphics[width=\linewidth]{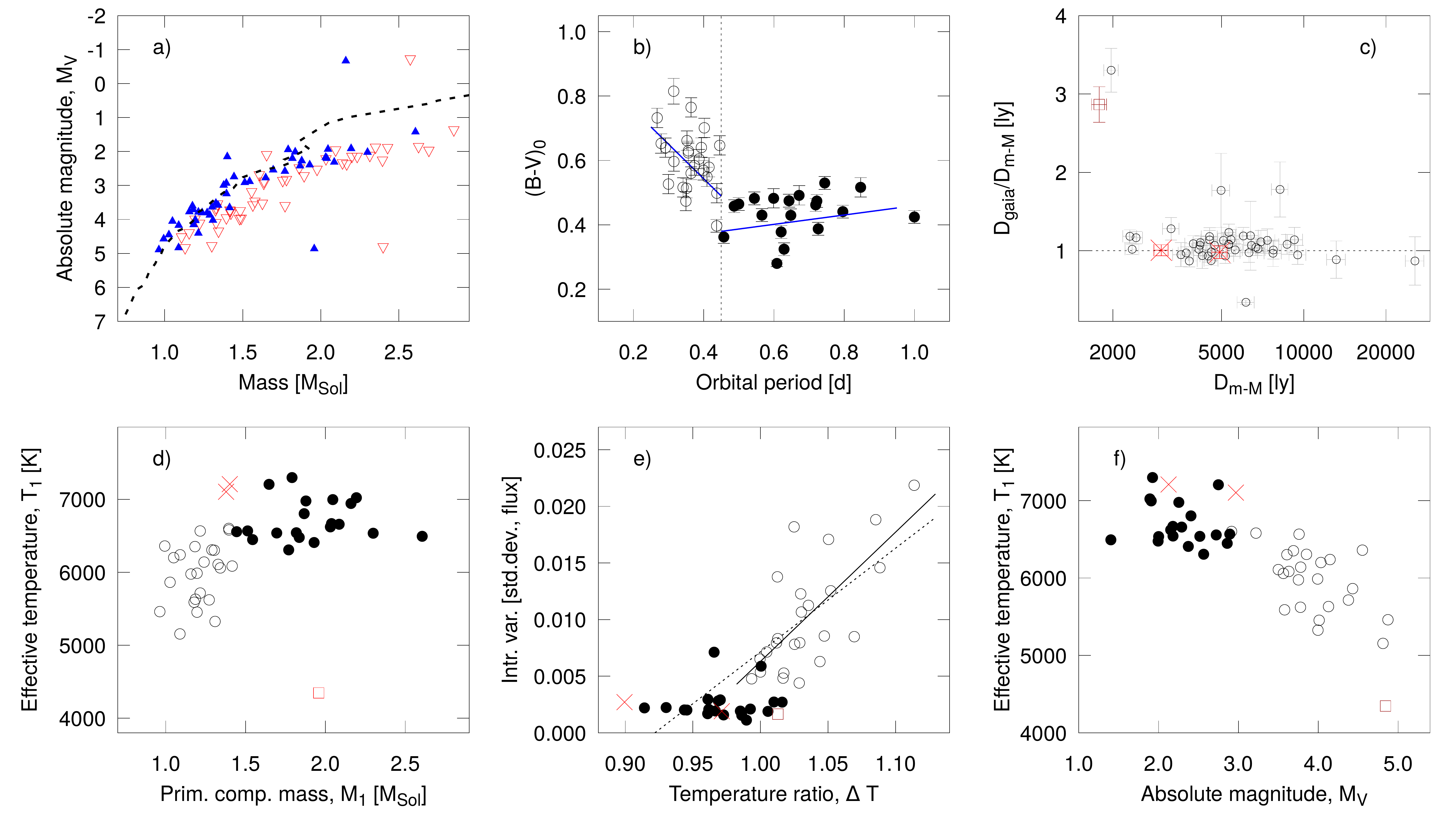}
    \caption{Relations between the observed and calculated parameters.  Panel a) shows the Mass - Absolute Magnitude relation for the total mass of a binary (empty triangles), and the primary star mass only (filled triangles). The dashed line shows the relation for single Main Sequence stars. Panel b) shows the Period-Color relation. The otherwise logarithmic \citet{eggen} relation can be successfully expressed as two linear relations, divided by the $P=0.45\,$d cut-off. Panel c) compares the distance modulus distances (D$_{m-M}$) with the distances calculated via the parallax ($D_{gaia}$). The red X-marks are the KIC 10267044 and KIC 5809868 systems. The brown square is the KIC 11618883. The remaining systems are shown as the black filled circles. Panel d) depicts the relation between the primary component mass and its effective temperature. Objects in binaries with $P<0.45$\,d are shown with empty circles, and the X-marks and empty square denote the same systems as in previous panel. Panel e) shows the relation between the intrinsic variability and the modeled temperature ratio of the binary. The scheme of the symbols is as in the previous panel. The dashed line is the linear fit to the $P<0.45$\,d systems; the solid line represent the same fit, but excluding the KIC 10267044 and KIC 5809868 systems. Panel f) shows the relation between the primary star effective temperature and the absolute magnitude of the system. Symbol scheme is as in the previous panel. The whole right row of this Figure presents Period-Color relation or its variations.}
    \label{fig:47corel}
\end{figure*}

\begin{table*}
    \centering
    \scriptsize
	\caption{Calculated masses of the binary components. The masses of primaries, $M_{1,G}$ were calculated with the \citet{2008MNRAS.390.1577G} formula. The masses of secondaries, $M_{2,G}$ were obtained by multiplying the primary masses by the photometric mass ratios established in this work. The $M_{1,r}$ and $M_{2,r}$ masses come from the total system masses calculated with Rucinski' absolute magnitudes. The $M_{1,g}$ and $M_{2,g}$ are calculated analogously basing on the Gaia EDR3 parallaxes. All masses are given is solar mass, $M_{\odot}$.}
	\label{tab:mass}
	\begin{tabular}{ccccccccccccccc}
	\hline
KIC \#      &$M_{1,G}$  &  $M_{2,G}$ &  $M_{1,r}$  &  $M_{2,r}$  &  $M_{1,g}$  &  $M_{2,g}$ & & KIC \#   &$M_{1,G}$  &  $M_{2,G}$ &  $M_{1,r}$  &  $M_{2,r}$  &  $M_{1,g}$  &  $M_{2,g}$   \\
\hline
2159783 & 1.240 &  0.180 &  0.951 &  0.138 &  1.229 &  0.179 & \, & 8682849 & 1.186 &  0.143 &  1.068 &  0.129 &  0.993 &  0.120 \\
2437038* & 0.963 &  0.168 &  0.713 &  0.122 &  4.013 &  0.688 & \, & 8804824 & 1.444 &  0.162 &  1.953 &  0.217 &  2.119 &  0.235 \\
2570289 & 0.994 &  0.114 &  0.452 &  0.052 &  0.393 &  0.045 & \, & 8842170 & 1.178 &  0.161 &  2.237 &  0.305 &  2.823 &  0.385 \\
3104113 & 2.299 &  0.393 &  1.382 &  0.242 &  2.470 &  0.432 & \, & 9030509 & 1.309 &  0.181 &  1.263 &  0.222 &  1.091 &  0.192 \\
3127873 & 1.929 &  0.210 &  1.547 &  0.169 &  1.301 &  0.142 & \, & 9087918 & 1.416 &  0.355 &  0.985 &  0.248 &  1.712 &  0.430 \\
3342425 & 1.289 &  0.135 &  0.604 &  0.063 &  0.954 &  0.100 & \, & 9151972* & 1.272 &  0.212 &  1.338 &  0.223 &  0.057 &  0.010 \\
4036687 & 1.050 &  0.144 &  0.891 &  0.122 &  0.688 &  0.094 & \, & 9283826* & 1.196 &  0.199 &  0.841 &  0.140 & 30.182 &  5.019 \\
4244929 & 1.158 &  0.168 &  1.267 &  0.183 &  1.618 &  0.234 & \, & 9350889 & 2.046 &  0.304 &  1.465 &  0.218 &  2.595 &  0.385 \\
5283839 & 1.090 &  0.134 &  0.671 &  0.082 &  0.991 &  0.122 & \, & 9453192 & 2.031 &  0.283 &  1.743 &  0.243 &  2.392 &  0.334 \\
5290305 & 1.818 &  0.381 &  2.564 &  0.540 &  2.868 &  0.604 & \, & 9703626 & 1.342 &  0.284 &  1.284 &  0.272 &  1.234 &  0.262 \\
5439790 & 2.194 &  0.432 &  1.498 &  0.295 &  1.590 &  0.313 & \, & 9776718 & 1.647 &  0.246 &  0.688 &  0.107 &  4.107 &  0.641 \\
5809868 & 1.401 &  0.252 &  3.140 &  0.454 &  3.186 &  0.461 & \, & 10007533 & 1.878 & 0.155 &  1.303 &  0.105 &  1.604 &  0.130 \\
6118779 & 1.216 &  0.129 &  0.616 &  0.065 &  0.776 &  0.082 & \, & 10229723* & 1.836 & 0.259 &  3.290 &  0.467 &  5.553 &  0.787 \\
7601767 & 1.513 &  0.243 &  1.543 &  0.248 &  2.606 &  0.420 & \, & 10267044 & 1.378 & 0.257 &  1.022 &  0.191 &  0.936 &  0.174 \\
7698650 & 1.770 &  0.206 &  1.731 &  0.201 &  2.433 &  0.282 & \, & 10322582 & 1.027 & 0.130 &  0.736 &  0.092 &  0.511 &  0.064 \\
7709086 & 1.328 &  0.250 &  1.323 &  0.249 &  1.824 &  0.344 & \, & 10395609 & 1.216 & 0.202 &  0.627 &  0.105 &  0.529 &  0.089 \\
7821450 & 1.089 &  0.213 &  0.960 &  0.188 &  1.896 &  0.371 & \, & 10528299 & 1.304 & 0.261 &  0.942 &  0.188 &  0.832 &  0.166 \\
8143757 & 1.196 &  0.284 &  1.787 &  0.423 &  1.720 &  0.407 & \, & 10618253 & 1.396 & 0.161 &  0.980 &  0.114 &  1.677 &  0.195 \\
8145477 & 1.695 &  0.167 &  1.670 &  0.165 &  2.198 &  0.217 & \, & 11097678 & 2.606 & 0.246 &  2.473 &  0.235 &  2.777 &  0.264 \\
8265951* & 2.160 &  0.413 & 58.599 &  9.059 & 37.043 &  5.727 & \, & 11144556 & 1.867 & 0.299 &  1.080 &  0.174 &  1.620 &  0.261 \\
8432859 & 1.183 &  0.146 &  0.930 &  0.115 &  0.648 &  0.080 & \, & 11618883* & 1.958 & 0.441 &  0.494 &  0.111 & 11.627 &  2.620 \\
8496820 & 1.395 &  0.241 &  1.522 &  0.263 &  1.550 &  0.268 & \, & 12055014 & 1.544 & 0.236 &  1.464 &  0.229 &  1.760 &  0.275 \\
8539720 & 2.086 &  0.309 &  1.161 &  0.172 &  1.542 &  0.228 & \, & 12352712 & 2.038 & 0.195 &  1.408 &  0.134 &  0.889 &  0.085 \\
8554005 & 1.791 &  0.637 &  1.625 &  0.579 &  1.780 &  0.634 & \, & & & & & & & \\

\hline
	\end{tabular}
\end{table*}

\section{The physical properties}
\label{sec:physical}

\subsection{The absolute magnitudes and distance calculation}
We started the exploration of physical parameters of the studied binaries by calculating the distances. Our primary method was to use the distance modulus. First, we transformed the magnitudes of the objects in \emph{g,r,i} filters \citep[][]{2011AJ....142..112B} into ($B-V$) colors and $V$ magnitudes in Johnson system using \citet{2005AJ....130..873J} formula. This was necessary to calculate the uncertainties, which were otherwise not provided in the catalogues. Then, we obtained the absolute magnitudes in $V$ filter using the \citet{2004NewAR..48..703R} relation. To calculate the dereddened colors and finally the extinction-corrected distance modulus, we used the \citet{2011ApJ...737..103S} extinction maps available at the NASA/IPAC Infrared Science Archive\footnote{\url{https://irsa.ipac.caltech.edu/applications/DUST/},\\accessed in November 2020}. We assumed the color excess follows the $A_V=3.1\,E(B-V)$ relation.

When reviewing the absolute magnitudes and the distances we found KIC 8265951 to clearly stand out from the remainder of the sample ($M_V$\,$=$\,$-0.69$\,mag). Because of the location of the binary close to a dense interstellar cloud, the abnormal absolute magnitude might have been an extinction-based error. To test this hypothesis we decided to verify all distances using the GAIA mission parallaxes \citep[Early Release DR3][]{gaia2021}. We adopted the EDR3\footnote{\url{https://gea.esac.esa.int/archive/}, accessed in June 2021} data and applied the mean offset for W UMa-type contact binaries \citep[][]{gaia_wuma2021}. As a result of the coordinate search in the GAIA EDR3, some of the binaries appeared as double sources. We dealt with that by choosing the objects with the closest magnitudes and colors to those in the KIC catalog, and therefore, in our numerical modeling.

The distances calculated in both methods, as well as the colors (not dereddened), $V$-band magnitudes and interstellar extinction $A_V$, are presented in Table~\ref{tab:dist}. For the full picture, we are presenting also the absolute magnitude calculated using the Rucinski's formula, $M_{V,r}$, as well as the absolute magnitudes obtained from the GAIA distances, $M_{V,g}$. Please note that the extinction we used is the total directional extinction, without prior estimation of the amount of dust between the observer and the object. This may have some impact on the absolute magnitude calculated with either method: the Rucinski's formula is color-dependent, while the GAIA distance absolute magnitude is extinction-dependent. It is worth noting that the absolute magnitudes obtained with GAIA data are on median 0.15~magnitudes lower (i.e. brighter) than their Rucinski's counterparts.

The distances obtained with the two methods are very consistent (Pearson correlation coefficient $r=0.9097$). The\,distance to KIC 8265951 was confirmed, which means that the system is located indeed on the outskirts of the dusty region Sh2-109. The discrepancies between the two methods of the distance determination is depicted in the Figure~\ref{fig:47corel}, panel c). Five objects are clearly standing out and need explanation.

First one, KIC 11618883, is marked as an empty square in the Figure 3. This is a system with the by far lowest effective temperature and the greatest ($B-V$) color. It has a barely resolved, white companion star. It is therefore highly probable, that the companion star is in fact the W UMa binary, while the red star is just a foreground object. We will verify this in an another study.

Next object, KIC 9151972, has a distance modulus $D_{m-M}$ overshoot. Most probably it is caused by the high third light contamination. We did not take the third light into account in any of the binaries, when calculating the distance modulus. The decision was dictated mainly by the unknown impact of the companion color on the (B-V) of the system.

The next two binaries, KIC 9283826 and KIC 2437038, stand out with their $\Delta\,T$ ratios, having a secondary star of a significantly greater apparent effective temperature than the primary. Their color might be under a large influence of their spot activity. It is possible that systems going through such an evolutionary stage might not follow the known Period-Color relation very well. Is should be noted though that those two binaries are not deviating from the Period-Color relation seen in Panel b) of the same Figure. Moreover, there is one system with an even greater temperature ratio: KIC 10322582. The latter binary is not experiencing distance discrepancy. Nevertheless, because its total eclipse is rather dubious, it might be the case that its inclination is in fact lower, the mass ratio greater and the resulting temperature ratio much closer to $\Delta\,T=1$. Such possibility should be investigated in an another study.

The last binary with a large distance difference is KIC 9776718. This system has a calculated color index ($B-V$)\,$=0.61$, although in the DSS maps it appears to have a very white color. We assume it is probable that the SDSS magnitudes we used for the ($B-V$) calculation might be flawed.

\subsection{Luminosity, radius and orbital separation}
A routine next step would involve calculation of the following physical parameters expressed in the solar units. The system total luminosity, $L_T$ can be derived from the absolute magnitude (equation~\ref{eq:lum}). The orbital separation, $A$, can be obtained using the system luminosity, effective temperatures of the components ($T_2$ from the numerical modeling) and the geometrical radii ($r_{1,2}$ constructed from the geometrical mean of the partial radii found via the numerical modeling, see equation~\ref{eq:separ}). Getting the stellar radii is a matter of multiplying the geometrical radii by the orbital separation. In addition, the total mas of the system, $M_{tot}$, can be found with the use of the Kepler law (equation~\ref{eq:mtot}). Individual masses of the components, $M_1$ and $M_2$, can be established by application of the mass ratio, $q$, from the light curve numerical modeling.
\begin{equation}
\label{eq:lum}
    L_{T}\, [L_{\odot}]=10^{-0.4(M_V-4.83)}
\end{equation} 
\begin{equation}
\label{eq:separ}
    A \,[R_{\odot}] = \sqrt{\frac{L_{T}}{T_1^4r_1^2+T_2^4r_2^2}}
\end{equation}
\begin{equation}
\label{eq:mtot}
    M_{tot} \,[M_{\odot}]=\frac{1}{74.53}\frac{A^3}{P^2}
\end{equation}
There are several issues with applying this chain of equations to the contact binaries. The effective temperature of the secondary (less massive) component can be overly exaggerated if the primary component has a highly spotted surface (this is explored further in Section~\ref{sec:deltem-intrvar}). From what we saw in Section~\ref{sec:intrvar}, virtually all contact binaries below the $P=0.45\,$d threshold are prone to a greater magnetic activity, and hence elevated probability of the spot presence. Because the median orbital period of all contact binaries around $P=0.375\,$d, that would mean the majority of the contact binaries have may have $T_2$ established erroneously high.

Another issue faced by the above calculations lies withing the partial radii found via numerical modeling. The numerical modeling of most of the contact binaries with a very low mass ratio tend to return very high fill-out factors. Moreover, in Section~\ref{sec:smaps} we have shown that both the averaged intrinsic variability and the phase smearing effects cause a very significant increase in the returned fill-our factors. That would cause an artificial increase of the geometrical radii by as much as 25\% (taking into account typical mass ratio of our sample). The combination of the overshoot in both radii and the secondary component effective temperature could mean an underestimation of the orbital separation by as much a s 30\%. More than that, the total luminosity of the system itself is highly uncertain in case of binaries with a third light. This is due to the high correlation between the third light, mass ratio and the system inclination.

This, in turn, translates to potential problems with the total masses of the systems. If the orbital separation is underestimated by as much as 30\%, then the masses could be erroneously lowered by as much as 60\%. This is obviously the worst case scenario, but it sets the boundaries of the possible estimation errors.

We calculated the radii and orbital separations for our systems, but we strongly advise to use them with caution. The table contains the geometrical radii, as well as the total luminosity and orbital separation calculated using two different absolute magnitudes: Rucinski's, $M_{V,r}$ and GAIA, $M_{V,g}$. The table is included in an appendix to this work (see Tab~\ref{tab:radii}).

\subsection{Calculation of the masses}

Because of the previously discussed issues, we decided to derive the masses of the studied binaries using another method. We chose to take advantage of the already existing \citet{2008MNRAS.390.1577G} Period-Mass relations for individual components. After the initial calculation we found that the relations produce mass ratios vastly different from our numerical modeling results. The differences were reaching as much as 100\%. We resolved that this is caused by a rather low correlation coefficient in the Period-Secondary Component Mass relation in \citet{2008MNRAS.390.1577G}. We dealt with this problem by refining the relation for the secondary component. The scheme was as following: first we derived the primary component masses using the original formula from their work:
\begin{equation}
\label{eq:m1}
     \text{log}\,M_1 = (0.755 \pm 0.059)\text{log}\,P + (0.416\pm 0.024)
\end{equation}
 
Then, we applied our photometric mass ratios to obtain the secondary component masses. Next, we fitted the such derived secondary component masses as a function of the orbital period with the log-log function. That allowed us to return the new and improved Period-Secondary Component Mass relation:
\begin{equation}
     \text{log}\,M_{2,\text{low}\,q} = (0.6907 \pm 0.1483)\text{log}\,P - (0.4049 \pm 0.0437)
\end{equation}

The discrepancy between the original relation for the secondary component masses and our solution is most probably due to the under-representation of the low-$q$ binaries in the \citet{2008MNRAS.390.1577G} work. We treat our solution as complementary to theirs, but in a low-$q$ regime. The calculated masses are shown in Table~\ref{tab:mass}. In the same table we show the masses calculated with the method discussed earlier. It is very problematic to estimate the true errors, and the formal uncertainties would most probably be underestimated. What is especially interesting, the GAIA-based total system masses are on median 23\% higher than their Rucinski's-based counterparts. Incidentally, the Gazeas-based total masses are on median 25\% higher than their Rucinski-based counterparts. Six objects marked with an asterisk in the table returned unrealistically high or low results for the latter two methods of the mass estimation. It is most likely due to the fact that in the GAIA EDR3 catalogue the values for those objects are different than for their counterparts in the KIC catalogue, despite identical coordinates.

\begin{table*}
\scriptsize
\centering
\renewcommand{\arraystretch}{1.6}
\setlength{\tabcolsep}{4pt}
	\caption{Calculated parameters of contact binaries in our sample. The $B-V$ colors and $V$ magnitudes were calculated with the \citet{2005AJ....130..873J} formula from the SDSS magnitudes. The interstellar extinction $A_V$ come from the \citet{2011ApJ...737..103S} reddening maps. The absolute magnitudes in $V$ filter were calculated with \citet{2004NewAR..48..703R} formula - $M_{V,r}$ - and with the use of the Gaia EDR3 parallaxes, $M_{V.g}$. The $D$ [ly] distances were calculated with the standard distance modulus method. The $D_{gaia}$ [ly] distanced were calculated from Gaia parallaxes for a verification.}
	\label{tab:dist}
	\begin{tabular}{cccccccrccccccccr}
	\hline
KIC \#       &$B-V$     & $V$       & $A_V$  & $M_{V,r}$ & $M_{V,g}$ &    $D$ [ly] & $D_{gaia}$ [ly] & & KIC \#       &$B-V$     & $V$       & $A_V$  & $M_{V,r}$ & $M_{V,g}$ &    $D_r$ [ly] & $D_{g}$ [ly]  \\
\hline
  2159783 &  0.77(3) & 15.1(2) & 0.56 & 3.78(9)  & 3.59(15)   & $  4711\pm 329$ & $  5039^{+ 160}_{- 150}$ & {} &   8682849 &  0.77(3) & 15.9(2) & 4.13(9)  & 4.18(27) & 4.18(27) & $  6300\pm 462$ & $  6147^{+ 347}_{- 312}$ \\ 
  2437038 &  0.87(3) & 16.2(2) & 0.42 & 4.87(9)  & 3.62(96)   & $  4980\pm 372$ & $  8816^{+2015}_{-1383}$ & {} &   8804824 &  0.41(2) & 14.7(1) & 2.72(6)  & 2.66(23) & 2.66(23) & $  7742\pm 526$ & $  7817^{+ 370}_{- 338}$ \\ 
  2570289 &  0.78(3) & 15.5(2) & 0.41 & 4.55(9)  & 4.65(51)   & $  4250\pm 304$ & $  3989^{+ 451}_{- 368}$ & {} &   8842170 &  0.83(3) & 15.1(2) & 3.58(9)  & 3.41(13) & 3.41(13) & $  3939\pm 274$ & $  4294^{+ 111}_{- 105}$ \\ 
  3104113 &  0.68(3) & 13.6(1) & 0.50 & 2.00(9)  & 1.58(12)   & $  5319\pm 333$ & $  6566^{+ 156}_{- 149}$ & {} &   9030509 &  0.87(3) & 14.7(1) & 4.00(9)  & 4.11(47) & 4.11(47) & $  3549\pm 240$ & $  3373^{+ 343}_{- 285}$ \\ 
  3127873 &  0.68(3) & 15.3(2) & 0.60 & 2.37(9)  & 2.50(29)   & $  9499\pm 670$ & $  8999^{+ 542}_{- 484}$ & {} &   9087918 &  0.74(3) & 14.6(1) & 3.63(9)  & 3.23(14) & 3.23(14) & $  4518\pm 305$ & $  5336^{+ 153}_{- 145}$ \\ 
  3342425 &  0.73(3) & 15.2(2) & 0.29 & 3.85(9)  & 3.52(21)   & $  5414\pm 380$ & $  6185^{+ 267}_{- 246}$ & {} &   9151972 &  0.73(3) & 15.5(2) & 3.78(9)  & 6.06(4)  & 6.06(4) & $  6140\pm 440$ & $  2104^{+  17}_{-  16}$ \\ 
  4036687 &  0.63(3) & 15.4(2) & 0.33 & 4.03(9)  & 4.22(16)   & $  5167\pm 366$ & $  4827^{+ 156}_{- 147}$ & {} &   9283826 &  0.72(3) & 13.2(1) & 3.99(9)  & 1.40(12) & 1.40(12) & $  1971\pm 120$ & $  6510^{+ 158}_{- 151}$ \\ 
  4244929 &  0.65(3) & 15.2(2) & 0.40 & 3.75(9)  & 3.57(19)   & $  5310\pm 372$ & $  5736^{+ 221}_{- 205}$ & {} &   9350889 &  0.52(2) & 13.6(1) & 1.91(6)  & 1.50(14) & 1.50(14) & $  6010\pm 378$ & $  7145^{+ 209}_{- 197}$ \\ 
  5283839 &  0.74(3) & 15.3(2) & 0.44 & 4.15(9)  & 3.87(18)   & $  4519\pm 319$ & $  5149^{+ 188}_{- 175}$ & {} &   9453192 &  0.51(2) & 14.1(1) & 2.15(6)  & 1.92(31) & 1.92(31) & $  7357\pm 477$ & $  8337^{+ 555}_{- 490}$ \\ 
  5290305 &  0.59(2) & 14.4(1) & 0.65 & 2.18(6)  & 2.10(16)   & $  6681\pm 443$ & $  6973^{+ 228}_{- 214}$ & {} &   9703626 &  0.65(3) & 15.6(2) & 3.56(9)  & 3.59(51) & 3.59(51)& $  7716\pm 556$ & $  7472^{+ 841}_{- 687}$ \\ 
  5439790 &  0.52(2) & 13.3(1) & 0.25 & 1.89(6)  & 1.85(9)    & $  5609\pm 344$ & $  5673^{+ 104}_{- 100}$ & {} &   9776718 &  0.61(2) & 15.1(2) & 2.75(6)  & 1.46(64) & 1.46(64) & $  8184\pm 571$ & $ 14586^{+2087}_{-1623}$ \\ 
  5809868 &  0.47(2) & 13.0(1) & 1.04 & 2.13(6)  & 2.12(5)    & $  3008\pm 180$ & $  3029^{+  29}_{-  28}$ & {} &  10007533 &  0.49(2) & 13.9(1) & 2.25(6)  & 2.10(12) & 2.10(12) & $  6432\pm 412$ & $  6876^{+ 167}_{- 159}$ \\ 
  6118779 &  0.85(3) & 15.7(2) & 0.25 & 4.38(9)  & 4.21(21)   & $  5314\pm 384$ & $  5764^{+ 250}_{- 230}$ & {} &  10229723 &  0.55(2) & 12.0(1) & 1.99(6)  & 1.61(6)  & 1.61(6) & $  2432\pm 135$ & $  2838^{+  36}_{-  35}$ \\ 
  7601767 &  0.54(2) & 14.6(1) & 0.25 & 2.89(6)  & 2.51(150)  & $  6381\pm 429$ & $  7605^{+2924}_{-1653}$ & {} &  10267044 &  0.47(2) & 14.1(1) & 2.97(6)  & 3.03(9)  & 3.03(9) & $  4923\pm 319$ & $  4836^{+  89}_{-  85}$ \\ 
  7698650 &  0.66(3) & 15.4(2) & 0.54 & 2.56(9)  & 2.31(34)   & $  9238\pm 654$ & $ 10532^{+ 775}_{- 676}$ & {} &  10322582 &  0.69(3) & 14.9(1) & 4.43(9)  & 4.69(8)  & 4.69(8) & $  3808\pm 262$ & $  3313^{+  57}_{-  55}$ \\ 
  7709086 &  0.82(3) & 16.0(2) & 0.85 & 3.50(9)  & 3.27(28)   & $  6965\pm 514$ & $  7758^{+ 465}_{- 415}$ & {} &  10395609 &  0.62(2) & 14.6(1) & 3.76(6)  & 3.88(9)  & 3.88(9) & $  4467\pm 301$ & $  4173^{+  75}_{-  72}$ \\ 
  7821450 &  0.92(4) & 15.2(2) & 0.34 & 4.81(12) & 4.32(21)   & $  3266\pm 228$ & $  4185^{+ 186}_{- 171}$ & {} &  10528299 &  0.63(3) & 14.1(1) & 3.61(9)  & 3.70(6)  & 3.70(6) & $  3710\pm 241$ & $  3608^{+  47}_{-  46}$ \\ 
  8143757 &  0.68(3) & 14.9(1) & 0.15 & 4.01(9)  & 4.04(22)   & $  4590\pm 315$ & $  4525^{+ 207}_{- 190}$ & {} &  10618253 &  0.64(3) & 12.9(1) & 3.22(9)  & 2.83(5)  & 2.83(5)& $  2309\pm 137$ & $  2735^{+  30}_{-  29}$ \\ 
  8145477 &  0.49(2) & 14.8(1) & 0.18 & 2.52(6)  & 2.32(24)   & $  8686\pm 594$ & $  9397^{+ 473}_{- 430}$ & {} &  11097678 &  0.53(2) & 13.3(1) & 1.40(6)  & 1.32(12) & 1.32(12) & $  6802\pm 418$ & $  6982^{+ 174}_{- 166}$ \\ 
  8265951 &  0.45(2) & 12.8(1) & 2.71 & -0.69(6) & -0.36(6)   & $  4580\pm 269$ & $  4007^{+  51}_{-  50}$ & {} &  11144556 &  0.56(2) & 13.6(1) & 2.41(6)  & 2.12(9)  & 2.12(9) & $  5082\pm 319$ & $  5754^{+ 104}_{- 101}$ \\ 
  8432859 &  0.67(3) & 17.2(2) & 0.48 & 3.69(9)  & 3.95(95)   & $ 13156\pm1043$ & $ 11667^{+2614}_{-1805}$ & {} &  11618883 &  1.41(6) & 13.8(1) & 4.84(18) & 2.55(8)  & 2.55(8) & $  1783\pm 113$ & $  5110^{+  82}_{-  79}$ \\ 
  8496820 &  0.55(2) & 12.7(1) & 0.47 & 2.91(6)  & 2.90(3)    & $  2353\pm 137$ & $  2398^{+  16}_{-  16}$ & {} &  12055014 &  0.53(2) & 13.6(1) & 2.86(6)  & 2.73(7)  & 2.73(7) & $  4168\pm 261$ & $  4466^{+  65}_{-  63}$ \\ 
  8539720 &  0.58(2) & 13.0(1) & 0.16 & 2.29(6)  & 2.08(8)    & $  4181\pm 250$ & $  4616^{+  78}_{-  76}$ & {} &  12352712 &  0.54(2) & 16.9(2) & 2.18(6)  & 2.51(138)& 2.51(138) & $ 25526\pm1984$ & $ 22215^{+7705}_{-4549}$ \\ 
  8554005 &  0.37(1) & 12.7(1) & 0.29 & 1.92(3)  & 1.85(9)    & $  4134\pm 243$ & $  4211^{+  73}_{-  71}$ & {} &			&		   &		 &		&		&	&			  &							 \\

\hline
	\end{tabular}
\end{table*}

\subsection{The Mass-Absolute Magnitude relation}
Contact binaries are known to be about one magnitude less bright than single main sequence stars of the same mass \citep[][]{2004NewAR..48..703R}. It is due to the more massive component being the primary source of energy shared across the common envelope. We have shown in panel a) of Fig.~\ref{fig:47corel} that this relation holds also in our dataset. Here the total mass of a binary, $M_{tot}$ comes from our modified Gazeas' method. The data is represented by the empty triangles in the plot. At the same time, despite the shared energy, the primary components follow the Mass-Absolute Magnitude relation to the point of reaching $M_1 \approx 1.9 M_{\odot}$ (filled triangles in the plot). In this relation we compare the masses with the absolute magnitude of a system as a whole, and we use the absolute masses from the Rucinski's method. The Mass-Absolute Magnitude relation for the single main sequence stars is taken from the \citet{2013ApJS..208....9P} updated by E.E.Mamajek\footnote{\url{https://www.pas.rochester.edu/~emamajek/EEM_dwarf_UBVIJHK_colors_Teff.txt}, version 2021.03.02}.

\subsection{The Period-Color relation}
Encouraged by the earlier findings on the light curve intrinsic variability, we constructed the Period-Color (PC) diagram for our sample. The relation is depicted in panel b) of Fig.~\ref{fig:47corel}. A traditional way of looking at the diagram is via the logarithmic relation \citep[][]{eggen}. We propose to look at the diagram through the lens of the $P=0.45$\,d limit. By doing so, two linear relations became apparent, instead of one logarithmic. Hence, the PC relation for $P<0.45$\,d binaries is:
\begin{equation}
     (B-V)_0 = (0.905\,\pm\,0.138) - (0.847\,\pm\,0.374)P.
\end{equation}
It must be noted that the correlation coefficient for this relation is rather weak: $r \approx 0.43$. For the $P>0.45$\,d binaries the PC relation is:
\begin{equation}
     (B-V)_0 = (0.367\,\pm\,0.078) + (0.105\,\pm\,0.116)P,
\end{equation}
with an even weaker correlation coefficient $r \approx 0.21$. Our sample contains objects with mass ratios mostly below $q<0.25$, therefore the double PC relation might not necessarily work in mass ratios above that limit. Nonetheless, it is tempting to say that the contact binaries are indeed divided into sub-populations based on the $P=0.45\,$d cut-off.

Another variant of the Period-Color relation is shown in panel d) of the Fig.~\ref{fig:47corel}. It is a relation between the Mass and Effective Temperature for the Primary Star. This relation makes it easier to see the two sub-populations in our sample. These are divided at $M_{cut}=1.43\,M_{\odot}$ and $T_{cut} \approx 6500\,K$. Please note the $M_{cut}$ comes directly from the Gazeas Period-Mass relation. Both sub-populations form clumps with no apparent trends. The effective temperatures come from the Kepler Input Catalog, and therefore have a rather large uncertainty (on average 400 K). Not counting the three outlying objects (mentioned earlier KIC 10267044, KIC 5809868, and KIC 11618883) the two populations stay well within their boundaries. The $P<0.45$\,d binaries have their primary component of masses $M_{1}<1.43\,M_{\odot}$ and effective temperatures $T_1<6500\,K$. The characteristics of the primary components of $P>0.45$\,d binaries are exactly opposite. 

Yet another version of the PC relation is shown in panel f) of the Fig.~\ref{fig:47corel}. This is the System Absolute Magnitude vs Primary Star Effective Temperature relation. Here the two sub-populations are even more separated, and therefore better visible. The absolute magnitude here comes from the Rucinski's method. 

In all panels: c), d), e) and f) we marked the outlying KIC~10267044 and KIC~5809868 systems ($P<0.45\,$d) with {X-marks} and the KIC 11618883 ($P>0.45\,$d) with the empty square.

\subsection{The relation between the Apparent Temperature Ratio and the Light Curve Intrinsic Variability}
\label{sec:deltem-intrvar}

The only parameter that we tied successfully to the magnitude of the light curve intrinsic variability is the apparent temperature ratio of a binary. The binaries with $P<0.45$\,d follow a weak trend of light curve intrinsic variability in a correlation with the apparent temperature ratio, $\Delta\,T=T_2/T_1$. The relation is shown in the Panel e) in Figure~\ref{fig:47corel} (the $P<0.45$\,d binaries are depicted with empty circles). If we neglect the KIC 10267044 and KIC 5809868 binaries (marked with X), then the relation is: 
\begin{equation}
    I_{ntr}V_{ar}=0.1142 \Delta T - 0.1079
\end{equation}
with the correlation coefficient $r=0.6971$. The $I_{ntr}V_{ar}$ is the magnitude of the light curve intrinsic variability as measured in Section~\ref{sec:intrvar}. If the two neglected binaries were to be taken into account, then the slope of the correlation would be $a=0.0915$ with the correlation coefficient $r=0.7162$. The one outlying binary from the $P>0.45$\,d population (filled circles) is KIC 8265951 - it's overshoot comes from the nature of the data treatment, and not from the intrinsic variability itself (see Appendix~\ref{appendix1} for more information).

\section{Conclusions}
\label{sec:conclusions}

\subsection*{The effect of averaged light curve intrinsic variability}

This work explored the impact of the averaged light curve intrinsic variability in contact binaries on the results of the light curve numerical modeling. We simulated the effect in low-$q$ contact binaries by averaging the synthetic longitudinal spot migration. We found that the intrinsic variability averaging leads to an increase of the apparent (i.e. resulting from the modeling) mass ratio. The apparent increase of $ \Delta q = 5\% $ was independent of the input mass ratio. We observe an increase of the apparent inclination and the apparent fill-out factor; for the latter parameter the increase occurs only for the lowest input mass ratio. The effective temperature of the secondary component remains unaffected.

The averaged intrinsic variability largely cancels itself out with the Kepler- and TESS-like phase smearing effect. When the two effects are taken into account, the apparent mass ratio is virtually unaffected for systems with an orbital period of $P \geq 0.35$\,d, while for systems with $P=0.3$\,d the apparent mass ratio is decreased by 3\% with respect to the input. The apparent mass ratio decrease was at non-negligible 9\% level for systems with a very short orbital period of $P=0.25$\,d. Both light curve-distorting effects strongly increase the apparent fill-out factor. The secondary component effective temperature remains unchanged in all cases.

\subsection*{Numerical modeling results}

We performed a light curve modeling of 47 totally eclipsing binaries from the KEBC database, mainly in search of the photometric mass ratio and the confirmation of their contact nature. We did not take into account the phase smearing nor the intrinsic variability averaging, because for majority of our objects these two effects neutralize each other in the mass ratio domain. Our modeling returned a contact configuration for all 47 systems. 21 of our objects were exhibiting the O'Connell effect in the phased and averaged 4-years-long light curve. 19 of them have a primary brightness maximum lower than the secondary. This can be interpreted as a dark region located on the trailing side of the binary. Numerical modeling returned an average longitude of such a long-lasting dark region to be $\lambda=302^{\circ}\,(\sigma=36^{\circ})$.

Next 17 objects in our sample were studied earlier by \citet{zola2017}, who took the phase smearing into account during the modeling. These objects have orbital periods $P>0.4$\,d. Our modeling returned a very similar mass ratios and temperature ratios to the previous study. Any deviations between our and their results bear no relation with the orbital period. In contrast to the previous study, we found it was not necessary to introduce a third light in three of the studied systems. This inclines us to conclude that taking the phase smearing into account during the numerical modeling of contact binaries with such a long orbital periods is unnecessary.

We resolved that out of 47 objects in our sample, no less than 53\% have a possible third companion. Three objects have a possible companion seen in the minima timing diagram, but their best fitting models show no third light.

We established that the distances to our objects calculated with the \citet{2004NewAR..48..703R} formula are in general in a good agreement with the GAIA EDR3 corrected parallaxes.

We refined the Orbital Period - Secondary Component Mass relation from \citet{2008MNRAS.390.1577G} to work with low-$q$ contact binaries.

Recently, \citet{2020PASJ...72..103L} published a study of 380 Kepler contact binaries, which partially includes our 47-object sample. Unfortunately, because of the parameter determination in their work we concluded that it is hardly possible to reliably compare our findings.

\subsection*{The $P=0.45\,$d cutoff}

Upon the examination of the magnitude of the intrinsic variability as a function of the orbital period, we found a sharp cutoff at $P$~$=$~$\,0.45$\,d. The light curves of objects above this limit show next to none measurable intrinsic variability. Below the cutoff the light curves immediately start to experience a severe intrinsic variability of various intensities. This means that the intrinsic averaging effect will not be present in contact binaries with orbital period $P>0.45$\,d. 

Following this cutoff, we found that the Orbital Period - Color relation can be expressed as two linear relations, instead of a logarithmic one. As a variant to this relation, we showed that the primary component Mass - Effective Temperature is populated by two distinct groups of objects, distinguished by the $P=0.45$\,d cutoff only. 

Most of the above observations coincide with the characteristics of the A- and W-subtypes of the W UMa-type contact binaries. The division itself became questionable in recent years, as more and more contact binaries were shown to switch in between the subtypes in rather short timescales. Our Period - Intrinsic Variability diagram (left panel in Fig.~\ref{fig:intravar}) provides a new perspective on the issue. Since the starspot-based intrinsic variability can distort all parts of the light curve, it is natural to expect that the $P<0.45$\,d binaries will be able to change their relative minima depths. Following this, the $P<0.45$\,d binaries can be either W- or A-subtype, depending on the moment of observations. On the other hand, the $P>0.45$\,d binaries remain always in the A-subtype. In this picture the division based on the relative minima depth is not reliable in the morphological sense. In our opinion the A/W-subtype division, if it is to stay, should remain in the purely phenomenological realm, i.e. as a description of a transient light curve, not the object itself. 

We infer that a proper distinction between the contact binary types should be based on the $P=0.45$\,d cutoff only. Granted, we expect this boundary might be violated to some extend by objects with a low enough inclination, but this would be the result of the projection effects of high-latitude, light curve-disturbing phenomena on the surface of the binary \citep[see][for a short introduction to the topic]{debski20}. A study of this phenomenon, along with the further investigation of the $P=0.45$\,d intrinsic variability boundary, should be presented in an upcoming paper, as it is far out of the scope of this work. At this point, we expect that the (partially) synchronous rotation reaches a critical velocity at the orbital period cutoff. With a shorter orbital period, the rotation causes the magnetic dynamo to cross an innate threshold and unbind itself. That would result in an immediate formation of large starspots occurring near the polar regions, just as it is in case of fast rotating single stars. Such scenario opens an interesting path toward studying the interiors of contact binaries.

\subsection*{Light curve intrinsic variability and the spot activity}

The objects with $P<0.45$\,d showed a wide range of intrinsic variability, while our simulations (Section~\ref{sec:intrvar}) were performed for a fixed-size starspot. Because of that, our simulations may not represent the situation occurring in all objects equally well. It would be best to estimate the influence of the light curve intrinsic variability averaging with some independent parameter. Unfortunately, the intensity of the intrinsic variability below $P=0.45$\,d is not correlated with the orbital period, effective temperature, nor the photometric color. The only parameter correlating with the intrinsic variability seems to be the apparent temperature ratio, $\Delta T$. It would be advisable to estimate a possible increase of the $\Delta q$ basing on the $\Delta\,T$.

Up to this point we only assumed that the apparent $T_2>T_1$ is a numerical artifact caused by the simplistic model, rather than a real temperature ratio. In this scenario, the apparent $T_2>T_1$ is induced by a presence of large, cool starspots on the surface of the primary star. The correlation of the apparent $\Delta\,T$ with the light curve intrinsic variability proves that the $T_2>T_1$ is tied to the starspot activity. Moreover, such effect would be magnified with a dark 'spot' residing very close or around the neck of the binary, on the primary component. Our findings about the 'long-prevailing spot' close to the binary neck (Section~\ref{sec:model-oconnell}) seems to back up such a possibility. On the other hand, the simulated light curve intrinsic variability (Section~\ref{sec:intrvar}) showed that the effect of averaging does not change the $T_2$. If the effect itself does not influence the $\Delta\,T$, then it acts as an indicator of the overall 'spottness' (the total spot coverage) of the primary component. The increased spot coverage on the primary would be therefore the main cause of the elevated levels of the apparent $T_2$.

\subsection*{The impact on the current and planned photometric missions}

We see that the results obtained in this work can be useful when studying contact binaries observed with a range a different surveys. The photometry from the Kepler, as well as TESS space telescopes, is expected to be burdened with both phase smearing and intrinsic variability averaging effects. On the other hand, the LSST and GAIA light curves, because of their short exposure times and long-time base, will result with no phase smearing and a rather significant intrinsic variability averaging effect. We expect the same effect to be of importance in the small telescope photometric surveys, such as SuperWASP, CRTS, LINEAR and many other.

\section*{Acknowledgements}

The case study of every object was done largely via the Simbad\footnote{\url{http://simbad.u-strasbg.fr/}} database \citep{simbad} and the Aladin software \citep{aladin1,aladin2}.

The Author gratefully acknowledges the financial support for this work given by the Polish National Science Centre under the Preludium 12 Grant no. 2016/23/N/ST9/01218. The Author would like to thank to his friends who made this work possible to finish, either by proofreading or by giving a moral support during last two exceptionally tough years. The Author expresses his gratitude to the reviewer of this work for the provided comments and suggestions which enriched the analysis and broadened the interpretation of the results.

\section{Data Availability}
The data underlying this article will be shared on reasonable request to the corresponding author. Some of the contact binary helpful formulas were integrated as free to use calculators on our project's webpage\footnote{\url{http://bade.space/soft/}}.




\bibliographystyle{mnras}
\bibliography{biblio} 



\appendix

\section{Measurement of the intrinsic variation}
\label{appendix1}

\begin{figure*}
    \centering
    \includegraphics[width=0.95\linewidth]{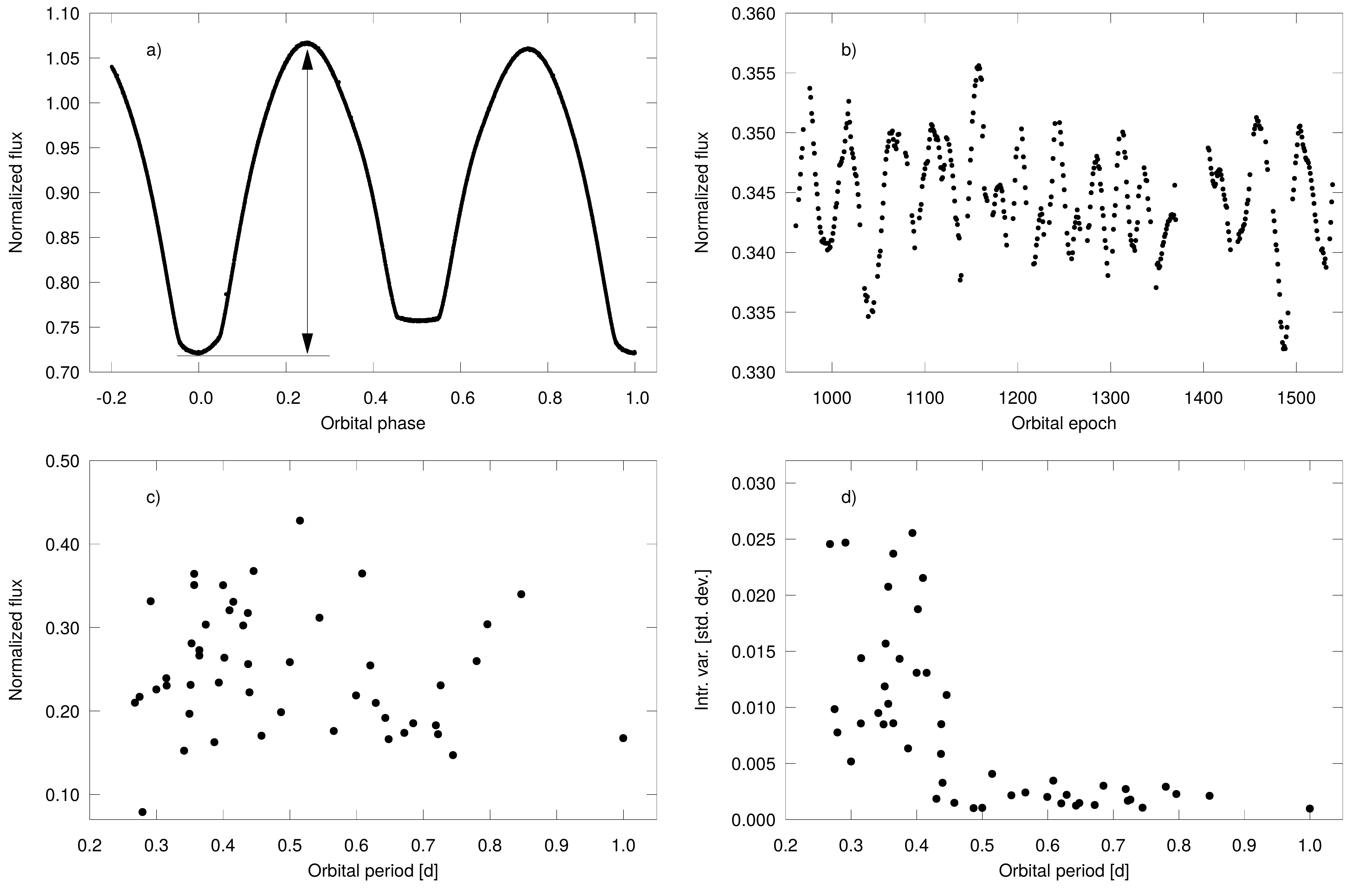}
    \caption{A visualization of an alternative method of the intrinsic variability quantification and measurement. For details see description in Appendix~\ref{appendix1}}.
    \label{fig:intrvar_alt}
\end{figure*}

In the main body of this work we adopted linear ephemeris for the phase-folding process. This choice is susceptible to the variability of the orbital period (real or apparent). Such variability can introduce an additional scatter of the phase-folded light curve. This period variation-caused scatter can easily be confused with the noise coming from the stellar intrinsic variability. It is especially visible in the left panel of Fig.~\ref{fig:intravar}, where one data point at $P=0.6$\,d stands out. This measurement comes from KIC\,8265951, which minima timing diagram shows clearly a strong light-time effect.

To confirm the $P=0.45$\,d sharp cutoff we measured the intrinsic variability in our sample with a range of alternative methods. One of such is based on a measurement of the standard deviation of the range between the height of the primary brightness maximum and the primary minimum. In the Figure~\ref{fig:intrvar_alt}, Panel a) depicts the measured light curve amplitude, and Panel b) shows the variation of the amplitude in terms of orbital epochs. Panel c) shows that there is no correlation between the mean amplitude of the light curve and the orbital period. Such correlation could influence the intrinsic variability vs orbital period relation measured with this particular method. The final results are shown in the Panel d) of Fig.~\ref{fig:intrvar_alt}. This diagram is very similar to the aforementioned Panel a) of Fig.~\ref{fig:intravar}, but this time the previously outlying data point from KIC\,8265951 is nicely following other $P>0.45$\,d systems. 

We reproduced the same sharp cut-off at $P=0.45$\,d with every method we used. In the main body of this work we chose to show the relation between the intrinsic variability and orbital period with the standard deviation of the phase-folded curve noise because this representation was the most straightforward. Other methods need explanation that is out of scope of this article and will be presented elsewhere.

\section{Radii, total luminosity and orbital separation}

\begin{table*}
    \centering
    \scriptsize
	\caption{Calculated stellar parameters: geometrical radii, $r_{1,2}$, total system luminosity, $L_{T,r}$, based on the Rucinski $M_V$, and $L_{T,g}$, based on GAIA $M_V$, and orbital separations $A_{r,g}$, calculated accordingly. The total luminosity is in units of solar luminosity, $L_{\odot}$, and the orbital reparations are in solar radii, $R_{\odot}$.}
	\label{tab:radii}
	\begin{tabular}{ccccccccccccccc}
	\hline
KIC \#      &$r_{1}$  &  $r_{2}$ &  $L_{T,r}$  &  $L_{T,g}$  &  A$_{r}$ [R$_{\odot}$]  &  $A_{g}$ [R$_{\odot}$] & & KIC \#      &$r_{1}$  &  $r_{2}$ &  $L_{T,r}$  &  $L_{T,g}$  &  $A_{r}$ [R$_{\odot}$]  &  $A_{g}$ [R$_{\odot}$] \\
\hline
   2159783 & 0.5742 & 0.2596  &   2.6303  &   3.1207  &   2.2470  &   2.4476 & \, &    8682849 & 0.5913 & 0.2519  &   1.9055  &   1.8151  &   2.2297  &   2.1761 \\
   2437038 & 0.5706 & 0.2867  &   0.9638  &   3.0486  &   1.6463  &   2.9280 & \, &    8804824 & 0.5949 & 0.2422  &   6.9823  &   7.3730  &   3.2345  &   3.3237 \\
   2570289 & 0.5985 & 0.2554  &   1.2942  &   1.1784  &   1.4300  &   1.3645 & \, &    8842170 & 0.5893 & 0.2829  &   3.1623  &   3.6922  &   2.8494  &   3.0789 \\
   3104113 & 0.5728 & 0.2960  &  13.5519  &  19.9603  &   4.4269  &   5.3726 & \, &    9030509 & 0.5758 & 0.3106  &   2.1478  &   1.9480  &   2.6143  &   2.4897 \\
   3127873 & 0.6038 & 0.2641  &   9.6383  &   8.5892  &   3.8634  &   3.6471 & \, &    9087918 & 0.5178 & 0.2831  &   3.0200  &   4.3651  &   2.6325  &   3.1650 \\
   3342425 & 0.6067 & 0.2610  &   2.4660  &   3.3438  &   1.9749  &   2.2997 & \, &    9151972 & 0.5788 & 0.3052  &   2.6303  &   0.3219  &   2.5916  &   0.9066 \\
   4036687 & 0.5893 & 0.2831  &   2.0893  &   1.7576  &   1.8934  &   1.7366 & \, &    9283826 & 0.5659 & 0.2722  &   2.1677  &  23.5900  &   2.1021  &   6.9346 \\
   4244929 & 0.5872 & 0.2910  &   2.7040  &   3.1826  &   2.3271  &   2.5247 & \, &    9350889 & 0.5815 & 0.2772  &  14.7231  &  21.5560  &   4.0427  &   4.8916 \\
   5283839 & 0.5937 & 0.2613  &   1.8707  &   2.4267  &   1.7737  &   2.0201 & \, &    9453192 & 0.5744 & 0.2521  &  11.8032  &  14.5742  &   4.2452  &   4.7173 \\
   5290305 & 0.5365 & 0.2748  &  11.4815  &  12.3710  &   4.4661  &   4.6359 & \, &    9703626 & 0.5331 & 0.2717  &   3.2211  &   3.1365  &   2.7140  &   2.6781 \\
   5439790 & 0.5400 & 0.2672  &  14.9968  &  15.6024  &   4.3916  &   4.4794 & \, &    9776718 & 0.5769 & 0.2786  &   6.7920  &  22.3582  &   2.5989  &   4.7153 \\
   5809868 & 0.5631 & 0.2429  &  12.0226  &  12.1385  &   3.7257  &   3.7436 & \, &   10007533 & 0.6105 & 0.2097  &  10.7647  &  12.3662  &   3.5321  &   3.7857 \\
   6118779 & 0.6038 & 0.2506  &   1.5136  &   1.7661  &   1.8885  &   2.0399 & \, &   10229723 & 0.5669 & 0.2446  &  13.6773  &  19.3905  &   4.8011  &   5.7166 \\
   7601767 & 0.5497 & 0.2445  &   5.9703  &   8.4677  &   3.1626  &   3.7664 & \, &   10267044 & 0.5511 & 0.2720  &   5.5463  &   5.2304  &   2.5564  &   2.4826 \\
   7698650 & 0.5917 & 0.2452  &   8.0910  &  10.1529  &   3.7248  &   4.1725 & \, &   10322582 & 0.5898 & 0.2565  &   1.4454  &   1.1324  &   1.7367  &   1.5372 \\
   7709086 & 0.5427 & 0.2630  &   3.4041  &   4.2176  &   2.6983  &   3.0034 & \, &   10395609 & 0.5780 & 0.3049  &   2.6792  &   2.3903  &   1.9346  &   1.8274 \\
   7821450 & 0.5402 & 0.2662  &   1.0186  &   1.6031  &   2.0391  &   2.5581 & \, &   10528299 & 0.5406 & 0.2705  &   3.0761  &   2.8320  &   2.3791  &   2.2828 \\
   8143757 & 0.5181 & 0.2731  &   2.1281  &   2.0740  &   2.7564  &   2.7211 & \, &   10618253 & 0.5949 & 0.2518  &   4.4055  &   6.3020  &   2.4991  &   2.9890 \\
   8145477 & 0.6028 & 0.2335  &   8.3946  &  10.0819  &   3.5245  &   3.8625 & \, &   11097678 & 0.6098 & 0.2398  &  23.5505  &  25.4404  &   5.8651  &   6.0959 \\
   8265951 & 0.5590 & 0.2492  & 161.4360  & 118.9060  &  14.5300  &  12.4701 & \, &   11144556 & 0.5802 & 0.3002  &   9.2897  &  12.1760  &   3.3801  &   3.8698 \\
   8432859 & 0.5937 & 0.2636  &   2.8576  &   2.2464  &   2.1262  &   1.8851 & \, &   11618883 & 0.5502 & 0.3071  &   0.9908  &   8.1359  &   2.7662  &   7.9266 \\
   8496820 & 0.5657 & 0.2793  &   5.8614  &   5.9329  &   2.9395  &   2.9573 & \, &   12055014 & 0.5712 & 0.2686  &   6.1376  &   6.9407  &   3.1587  &   3.3590 \\
   8539720 & 0.5819 & 0.2773  &  10.3753  &  12.5342  &   3.8045  &   4.1816 & \, &   12352712 & 0.6088 & 0.2376  &  11.4815  &   8.4500  &   3.9132  &   3.3571 \\
   8554005 & 0.5071 & 0.3335  &  14.5881  &  15.5002  &   3.9322  &   4.0533 & \, & & & & & & & \\
\hline
	\end{tabular}
\end{table*}

\section{The light curves and the best fitting models}

\begin{figure*}
	\includegraphics[width=\linewidth]{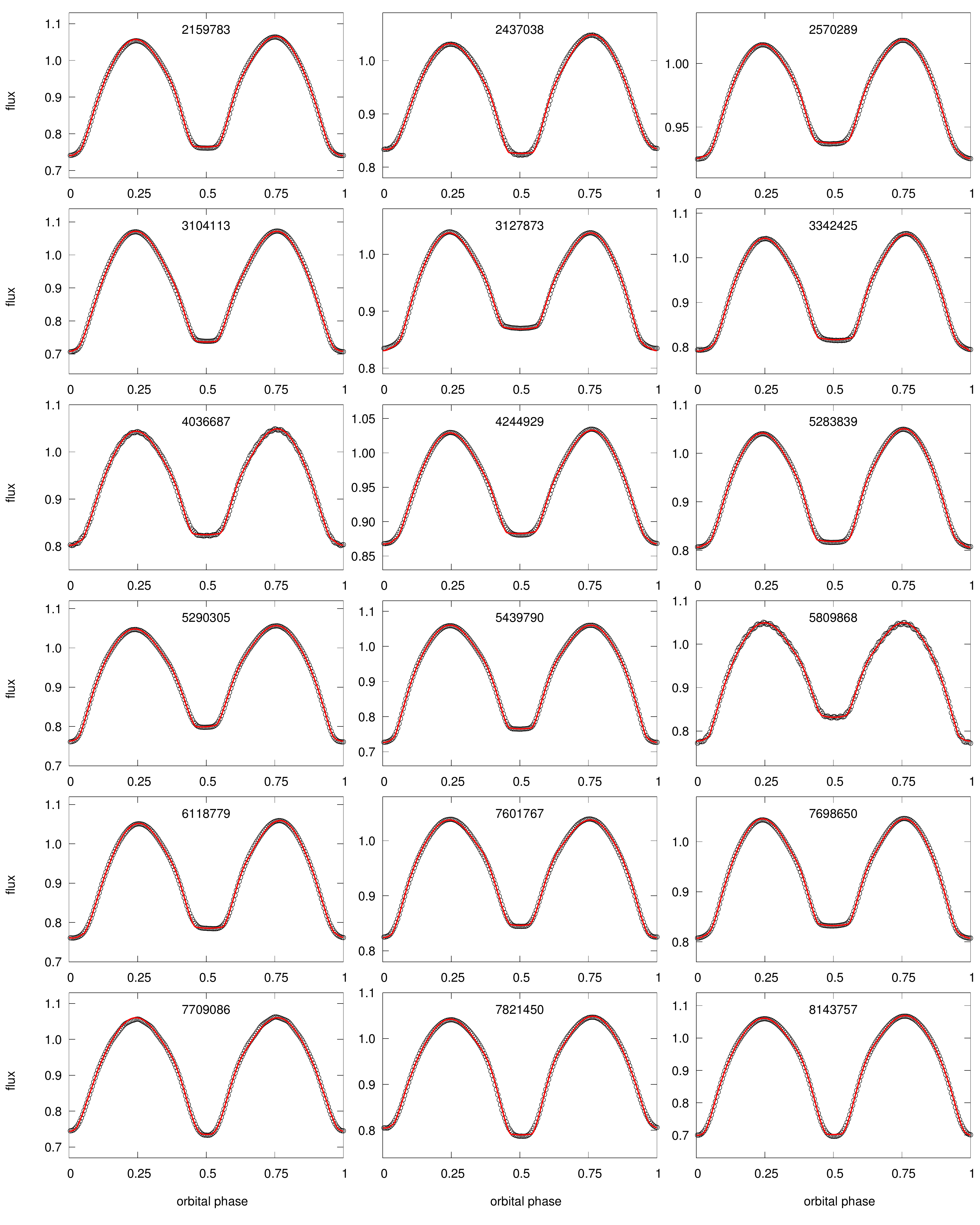}
    \caption{light curves of the contact binaries studied in this work (open circles) superimposed with the best fitting models (red line). The number in each panel is the KIC number for a given object.}
    \label{fig:lc-1}
\end{figure*}

\begin{figure*}
	\includegraphics[width=\linewidth]{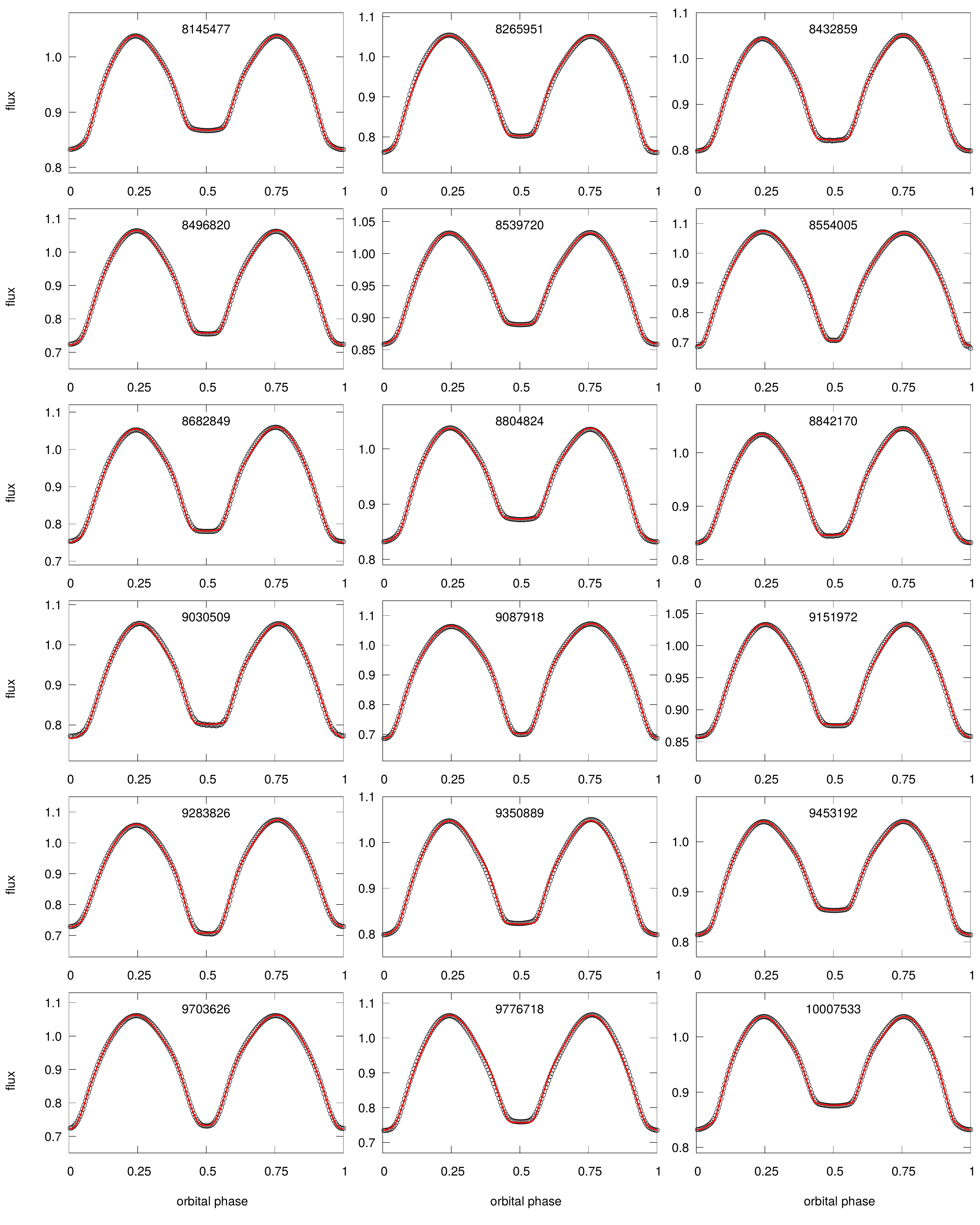}
    \caption{light curves of the contact binaries studied in this work (open circles) superimposed with the best fitting models (red line). The number in each panel is the KIC number for a given object. Continuation of the Fig.~\ref{fig:lc-1}.}
    \label{fig:lc-2}
\end{figure*}

\begin{figure*}
	\includegraphics[width=\linewidth]{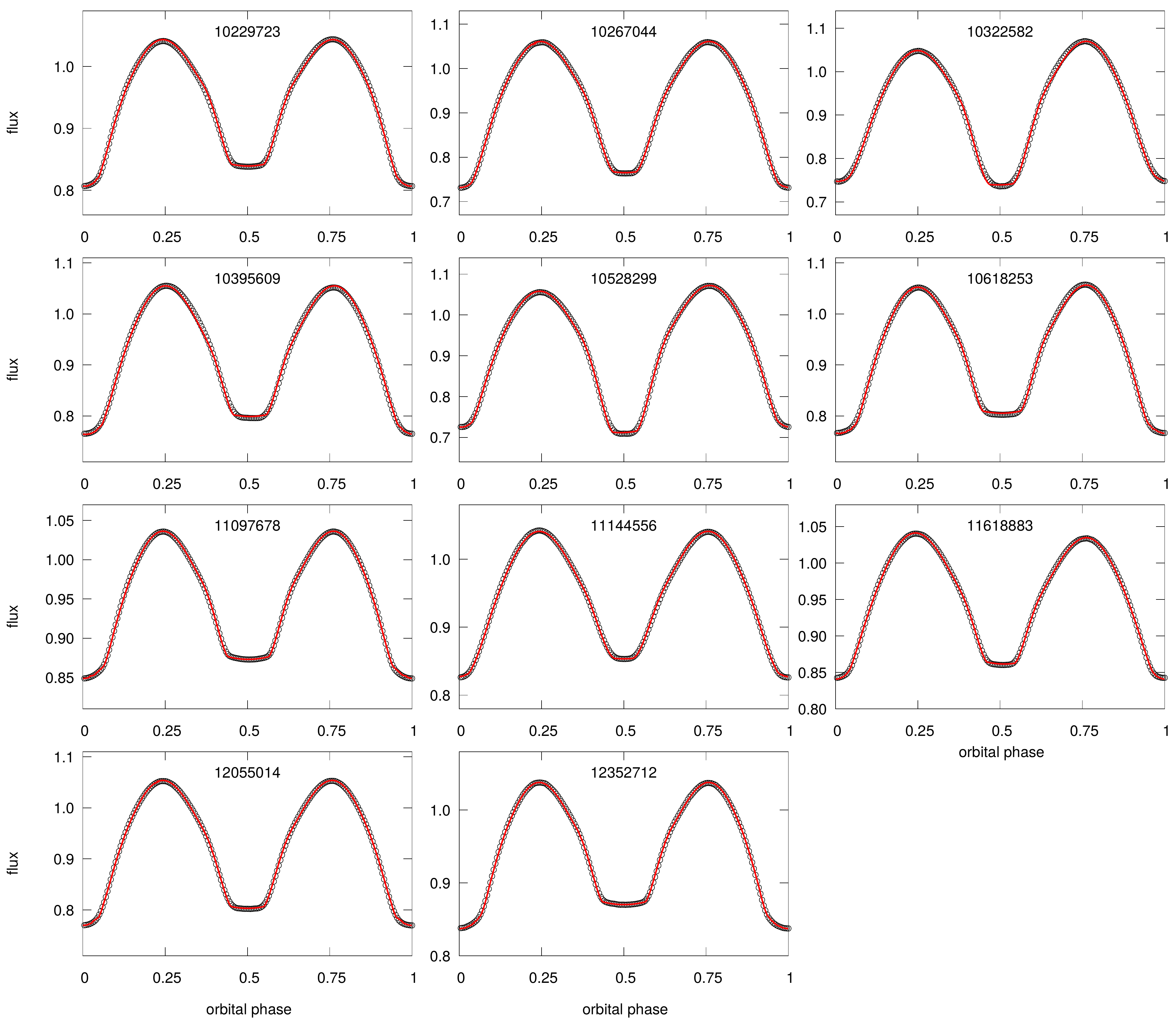}
    \caption{light curves of the contact binaries studied in this work (open circles) superimposed with the best fitting models (red line). The number in each panel is the KIC number for a given object. Continuation of the Fig.~\ref{fig:lc-1}.}
    \label{fig:lc-3}
\end{figure*}

Figures~\ref{fig:lc-1} to~\ref{fig:lc-3} contain phased and averaged light curves of the 47 studied objects. The light curves are shown as empty circles. The best-fitting models are superimposed over the light curves as a continuous line. Each panel is described by the KIC number of the given object's light curve. the light curves are presented in the units of the normalized flux. 

\bsp	
\label{lastpage}
\end{document}